\documentclass[prb, aps,amssymb, amsmath, twocolumn, superscriptaddress]{revtex4-2}

\usepackage{braket,bm,graphicx}
\usepackage{xcolor}

\begin{document}


\title{Weyl superconductivity and quasiperiodic Majorana arcs in quasicrystals}



\author{Masahiro Hori}
\email{mhorijapan@gmail.com}
\affiliation{%
Department of Physics and Engineering Physics, and Centre for Quantum Topology and Its Applications (quanTA), University of Saskatchewan, 116 Science Place, Saskatoon, Saskatchewan, S7N 5E2, Canada
}%
\affiliation{%
Department of Applied Physics, Tokyo University of Science, Tokyo 125-8585, Japan
}%

\author{Ryo Okugawa}
\affiliation{%
Department of Applied Physics, Tokyo University of Science, Tokyo 125-8585, Japan
}%

\author{K. Tanaka}
\affiliation{%
Department of Physics and Engineering Physics, and Centre for Quantum Topology and Its Applications (quanTA), University of Saskatchewan, 116 Science Place, Saskatoon, Saskatchewan, Canada S7N 5E2
}%

\author{Takami Tohyama}
\affiliation{%
Department of Applied Physics, Tokyo University of Science, Tokyo 125-8585, Japan
}%


\date{\today}

\begin{abstract}
    Weyl superconductivity is a topological phase in three-dimensional crystals in which the Weyl equation describes quasiparticle excitation near band-touching points in momentum space called Weyl nodes.
    For quasicrystals which lack translational symmetry, a theory of Weyl superconductivity has not been established, in spite of recent extensive studies on quasicrystalline topological phases.
    Here, we demonstrate the occurrence of quasicrystalline Weyl superconductivity by extending the definition of Weyl superconductivity to periodically stacked, two-dimensional superconducting quasicrystals. 
    We identify quasicrystalline Weyl nodes -- topologically protected point nodes in one-dimensional momentum space corresponding to the stacking direction -- in terms of a topological invariant given by a change in the Bott index in quasicrystalline layers.
    We find that these Weyl nodes exist in pairs and that Majorana zero-energy modes protected by the nonzero Bott index between a pair of quasicrystalline Weyl nodes appear on surfaces. 
    These Majorana zero modes form an infinite number of arcs in momentum space, densely and quasiperiodically distributed as a function of momentum in the direction of surfaces within each quasicrystalline layer.
    In Ammann-Beenker (Penrose) quasicrystals, the quasiperiodicity of Majorana arcs is governed by the silver (golden) ratio associated with the quasicrystalline structure.
\end{abstract}


\maketitle

\section{Introduction}

Quasicrystals (QCs) are materials whose structure is aperiodic with a long-range order \cite{Shechtman_1984}.
Recent experiments in QCs have shown the presence of electronic long-range orders, similarly to crystals, such as a long-range magnetic order \cite{Tamura_2021} and superconductivity \cite{Kamiya_2018, Tokumoto23}. On the other hand, it has been shown theoretically 
that quasicrystalline superconductors exhibit Cooper pairs with finite center-of-mass momentum \cite{Sakai_2017, Sakai_2019, Takemori_2020, Fukushima_2023} and anomalous paramagnetic response \cite{Fukushima_2023_2}, which are different from crystalline superconductors. Furthermore, some 
theoretical works have predicted topological phases in QCs without crystalline counterparts \cite{Fan21}.

One cannot use a topological band theory for QCs because 
they lack translational symmetry.
Nevertheless, in quasicrystalline systems, topological insulator 
phases beyond a topological band theory \cite{Tran15, Bandres16, Huang18PRL, Huang18PRB, Huang19, He19, Varjas19, Chen20, Duncan20, Hua20}, topological charge pumping \cite{Kraus12, Nakajima21, Yoshii_2021, Koshino22, Yamamoto22}, and topological semimetals \cite{Timusk13, Pixley18, Mastropietro20, JeffreyPNAS20, Grossi23, Mao23arXiv, Chen23arXiv} have been proposed theoretically.
In particular, layered QCs have attracted attention recently as a platform for realizing topological semimetallic phases \cite{JeffreyPNAS20,Grossi23,Mao23arXiv,Chen23arXiv}.

Some crystalline topological superconductors show gapless points in momentum space called point nodes.
If those points cannot be gapped out by weak perturbations that conserve symmetries of the system,
such points are called topologically protected nodes.
For example, a Weyl superconductor (WSC) is a three-dimensional (3D) superconductor with topologically protected point nodes called Weyl nodes, which exist pairwise and are described by the Weyl equation with chirality $\pm 1$ \cite{Meng12, Sau12, Fischer14, Yang14, Bendik_2015, Yuan17, Li_2018, Okugawa18, Sumita18, Nakai20}. In 3D momentum space, the Chern number changes at
each Weyl node by its chirality \cite{Wan11,Burkov11,Okugawa14}.
WSCs also have zero-energy surface states 
protected by the nonzero Chern number 
between a pair of Weyl nodes 
\cite{Schnyder15}. 
Those zero-energy modes are localized at the surfaces and form arcs in momentum space.
These arcs are called Majorana arcs because zero-energy modes in topological superconductors 
can be described as Majorana fermions \cite{Read_2000}.
Also in QCs, some theoretical works \cite{Fulga16, Cao20, Wang22, Liu23} including studies by some of the present authors \cite{Ghadimi21, Hori24} have proposed topological superconductivity with no node. 
Whether a quasicrystalline topological superconductor with nodes can exist 
or not has yet to be understood
because the nodes generally appear in the Brillouin zone 
of periodic systems. 

In this work, we 
study periodically stacked, two-dimensional (2D) quasicrystalline topological superconductors.
We find that this system shows topologically protected nodes in one-dimensional momentum space corresponding to the stacking direction. 
At these nodes, the Bott index 
changes its value in the same way as does the Chern number at Weyl nodes in crystalline WSCs. 
We thus call them quasicrystalline Weyl nodes \cite{Grossi23} and the resulting superconductors quasicrystalline WSCs.
Between two of such quasicrystalline Weyl nodes in momentum space, zero-energy modes protected by the 
nonzero Bott index appear on surfaces. 
Fourier momenta at which the Fourier amplitudes of these zero-energy surface modes are finite are densely and quasiperiodically distributed in one-dimensional momentum space that corresponds to the direction of surfaces within each layer. 
This quasiperiodic distribution of momenta of the motion along each surface in 2D QCs is in stark contrast to periodic distribution due to repeating Brillouin zones 
of layered 2D crystalline topological superconductors.

The paper is organized as follows.
We first review a 2D quasicrystalline topological superconductor in Sec.~\ref{sec:model_2D} and then introduce a layered quasicrystalline superconductor in Sec.~\ref{sec:model_3D}.
In Sec.~\ref{sec:results_quasicrystallineWeylNodes}, we extend the definition of Weyl superconductivity in crystals to QCs and obtain a topological phase diagram for a layered quasicrystalline Weyl superconductor.
To understand the difference between quasicrystalline and crystalline WSCs, we calculate the spectral density of states (SDOS) for topologically protected zero-energy modes in Sec.~\ref{sec:results_majoranaArcs}.
Finally, we summarize our work 
in Sec~\ref{sec:summary}.

\section{Model}
\label{sec:model}
\subsection{Two-dimensional quasicrystalline superconductor}
\label{sec:model_2D}
We consider 2D $s$-wave topological superconductivity with Rashba spin-orbit coupling in magnetic field perpendicular to the 2D ($xy$) plane \cite{Sato09,Sato10} in a QC, which is modeled as \cite{Ghadimi21}
\begin{align}
    H^{{\rm 2D}}=\frac{1}{2}\sum _{r r^{\prime}\sigma \sigma ' }
    (c_{r\sigma }^{\dagger}~ c_{r\sigma })
    \begin{pmatrix}
        \mathcal{H}_0^{\rm 2D} (\mu) & \Delta \\
        {\Delta}^{\dagger} & -\mathcal{H}_0^{\rm 2D} (\mu)^*
    \end{pmatrix}
    \begin{pmatrix}
        c_{r^{\prime}\sigma ' } \\
        c_{r^{\prime}\sigma ' }^{\dagger}
    \end{pmatrix},        
    \label{BdG}
\end{align} 
where $c_{r\sigma}^{\dagger}$ is the creation operator of an electron with spin $\sigma(=\uparrow,\downarrow)$ at the $r$th vertex. We use the vertex model, where the vertices of the quasiperiodic tiling are the lattice sites of a quasicrystal and the electron moves from a vertex to another vertex along the edge of a tile.

The 2D normal-state Hamiltonian $\mathcal{H}_0^{\rm 2D} (\mu)$ is given by
\begin{align}
\begin{split}
    &[\mathcal{H}_0^{\rm 2D} (\mu)]_{r\sigma, r^{\prime}\sigma '}\\
    &=[(t_{r r^{\prime}}-\mu \delta _{r r^{\prime}})\sigma _0
    -h_z\delta _{r r^{\prime}}\sigma _z
    +i\lambda _{r r^{\prime}}\bm{e}_z \cdot \bm{\sigma} \times \hat{R}_{r r^{\prime}}]_{\sigma \sigma '},
\end{split}
\label{eq:normal_state_Ham2D}
\end{align}
where $t_{r r^{\prime}}\equiv -t$ is the hopping amplitude along nearest-neighbor links between the $r$th and $r'$th vertices, $\mu$ the chemical potential, $-h_z$ ($+h_z$) the Zeeman energy due to the magnetic field for spin up (down), $\lambda _{r r^{\prime}} \equiv \lambda$ the Rashba spin-orbit coupling constant along nearest-neighbor links between the $r$th and $r'$th vertices, $\bm{e}_z$ a unit vector in the direction perpendicular to the plane, $\bm{\sigma}=(\sigma _x,\sigma _y, \sigma _z)$ the Pauli matrices acting on spin space, $\sigma _0$ an identity matrix, and $\hat{R}_{r r^{\prime}}$ the normalized vector from the $r$th to $r'$th vertices. In the following, we set $t$ as the unit of energy. An $s$-wave superconducting pairing operator $\Delta$ is defined as
\begin{align}
    [{\Delta} ]_{r\sigma ,r^{\prime}\sigma '}=[\delta _{r r^{\prime}} i \Delta _0\sigma _y]_{\sigma \sigma '},
\end{align}
where $\Delta _0$ is the superconducting order parameter. In the following, we set $\lambda =0.5t$, $h_z=-0.5t$, and $\Delta _0=0.2t$. 

\begin{figure}
	\includegraphics[width=8.6cm]{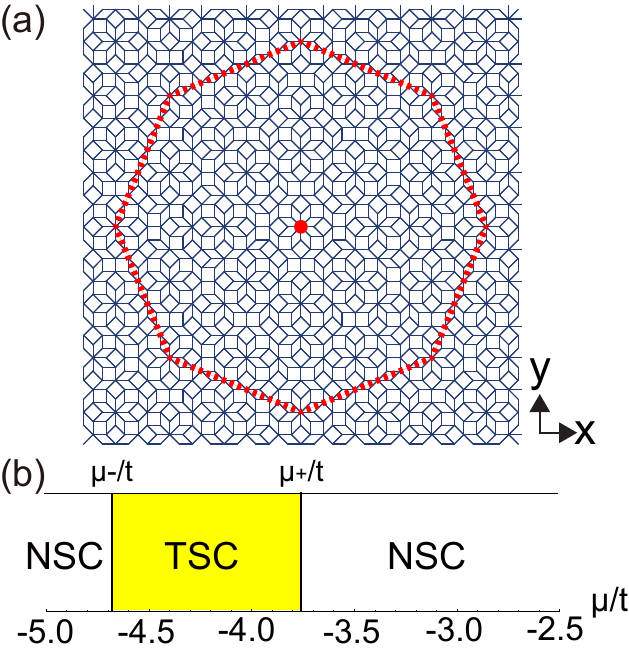}
	\caption{\label{AB} 
        (a) Ammann-Beenker approximant of square shape with 1393 vertices. The structure is identical to that of an Ammann-Beenker QC inside the octagon outlined by red dashed lines. The red circle indicates the center of the Ammann-Beenker QC.
        (b) Topological phase diagram as a function of chemical potential $\mu$ of the 2D model in Eq.~(\ref{BdG}) on the Ammann-Beenker QC.
        TSC (NSC) represents topological (normal, i.e., trivial) superconductivity with $B=1$ ($B=0$). 
        }
\end{figure}

The Hamiltonian Eq.~(\ref{BdG}) describes a topological superconductor with time-reversal symmetry broken by the magnetic field, and this model belongs to class $D$ in the Altland-Zirnbauer classification \cite{Altland_1997}.
In 2D systems of class $D$, it is possible to define a topological invariant that takes on integer values.
The topological invariant changes its value only when the bulk energy gap vanishes, and Majorana edge modes can appear when the invariant is nonzero in the bulk \cite{Schnyder_2008, Chiu_2016}.
Among such topological invariants, the Bott index $B$ is applicable not only for periodic systems, but also for aperiodic systems including QCs (see Appendix \ref{Bott}) \cite{Loring10, Ghadimi21}.
The Bott index for $H^{{\rm 2D}}$ in Eq.~(\ref{BdG}) can vary as a function of $\mu$, for example.
Figure~\ref{AB}(a) illustrates an Ammann-Beenker approximant of square shape that contains 1393 vertices. Approximants are identical to QCs except for regions close to the edges, to which the periodic boundary condition (PBC) can be applied.
In Fig.~\ref{AB}(a), vertices enclosed by the octagon in red dashed lines compose an Ammann-Beenker QC.
Since calculation of the Bott index requires PBC 
imposed on the system, we
use an Ammann-Beenker approximant to study the topological properties of Ammann-Beenker QCs.
In Appendix~\ref{Penrose} we show our study of the topological properties of Penrose QCs in terms of a Penrose approximant.
Figure~\ref{AB}(b) is a topological phase diagram for a superconducting Ammann-Beenker QC for low electron density \cite{Ghadimi21}. A topological phase transition occurs when $B$ changes due to closing of the bulk spectral gap. In Fig.~\ref{AB}(b), $\mu _{+}/t \simeq -3.76$ and $\mu _{-}/t \simeq -4.68$ are the chemical potential at the topological phase boundaries \cite{Ghadimi21}, such that 
$B=1$ for $\mu _{-} < \mu < \mu _{+}$ and $B=0$ for $\mu$ outside this region. 

\subsection{Layered quasicrystalline superconductor}
\label{sec:model_3D}
To define quasicrystalline Weyl superconductivity, we consider periodic stacking of infinitely many, 2D quasicrystalline topological superconductors described in Sec.~\ref{sec:model_2D}; i.e.,  
with magnetic field in the stacking ($z$) direction 
and intralayer Rashba spin-orbit coupling.
As a 2D quasicrystalline system in the $xy$ plane, we use Ammann-Beenker QCs and apply PBC in the $z$ direction.
We introduce interlayer nearest-neighbor hopping:
\begin{align}
    H_z=\sum _{k_z}2t_z\cos (k_zc)c^{\dagger}_{k_{z}r\sigma }c_{k_{z}r\sigma },
    \label{Hz}
\end{align}
where $k_z$ and $c$ are, respectively, momentum and the interlayer distance in the $z$ direction, $t_z$ is the interlayer hopping amplitude, and $c_{k_{z}r\sigma }^{\dagger}$ is the creation operator of an electron at the $r$th vertex in each layer with spin $\sigma(=\uparrow, \downarrow)$ and momentum $k_z$. Since there is an infinite number of layers, $k_z$ is a continuous variable in the Brillouin zone, $-\pi/c \le k_z < \pi/c$.
Thus, a layered quasicrystalline superconductor is described by
\begin{align}
\begin{split}
    H^{\mathrm{3D}}=\frac{1}{2}\sum _{k_z}\sum _{r r^{\prime}\sigma \sigma ' }&
    (c_{k_{z}r\sigma }^{\dagger}~ c_{k_{z}r\sigma })\\
    &\begin{pmatrix}
        \mathcal{H}_0^{\mathrm{3D}}(k_z) & \Delta \\
        {\Delta}^{\dagger} & -{\mathcal{H}_0^{\mathrm{3D}}(k_z)}^*
    \end{pmatrix}
    \begin{pmatrix}
        c_{k_{z}r^{\prime}\sigma ' } \\
        c_{k_{z}r^{\prime}\sigma ' }^{\dagger}
    \end{pmatrix}.        \label{3DBdG}
\end{split}
\end{align}
Here the 3D normal-state Hamiltonian is given by $[\mathcal{H}_0^{\mathrm{3D}}(k_z)]_{r,r'}=[\mathcal{H}_0 ^{\rm 2D}(\mu)]_{r, r'}+2t_z \cos{(k_z c)} \delta_{rr'} \sigma_0$, which can be written as
\begin{align}
    \mathcal{H}_0^{\mathrm{3D}}(k_z)=\mathcal{H}_0 ^{\rm 2D}(\tilde{\mu}(k_z))
    \label{3DHam}
\end{align}
with $\tilde{\mu}(k_z) = \mu -2t_z\cos k_z c$. Therefore, $\tilde{\mu}(k_z)$ can be regarded as a $k_z$-dependent effective chemical potential in the 2D Hamiltonian.

A possible setup for realizing our model is alternating layers of a 2D superconducting quasicrystal and a ferromagnetic insulator. This is analogous to a 2D quasicrystal sandwiched between a superconductor and 
a ferromagnetic insulator proposed in Ref.~\cite{Ghadimi21}, and also to the experimental realization of 2D topological superconductivity in monolayer lead covering a cobalt island \cite{Menard_2017}. Another possible approach is doping layered superconducting quasicrystals with magnetic impurities. For periodic systems, WSC has been suggested to occur in superlattices of superconductors and magnetically doped topological insulators \cite{Meng12}. 

\section{Results}
\label{sec:results}
\subsection{Quasicrystalline Weyl nodes}
\label{sec:results_quasicrystallineWeylNodes}
We introduce a $k_z$-dependent Bott index $B(k_z)$ for a given $k_z$ in 
RHS of Eq.~(\ref{3DBdG}) and 
define a topological invariant $\chi$ at $k_z=k_0$ as
\begin{align}
    \chi (k_0)= \lim _{\delta \rightarrow 0^+}[B(k_0+\delta)-B(k_0-\delta)].
    \label{eq:chirality}
\end{align}
Nonzero $\chi (k_0)$ implies closing of the spectral gap as the Bott index changes at $k_z=k_0$. 
We call such a nodal point in momentum space 
a quasicrystalline Weyl node \cite{Grossi23} 
and define superconductivity with such Weyl nodes in QCs as quasicrystalline Weyl superconductivity.

For a given $k_z$, the 3D Hamiltonian of our model 
contains $H_0^{{\rm 2D}}(\tilde{\mu}(k_z))$ via Eq.~(\ref{3DHam}). 
This means that if there is a quasicrystalline Weyl node at $k_z=k_0$, $\tilde{\mu}(k_0)$ corresponds to the chemical potential at which the 3D Hamiltonian for a given $k_z=k_0$ exhibits a topological phase transition, as illustrated in Fig.~\ref{AB}(b). That is, $|k_0|$ is equal to either $|k_+|$ or $|k_-|$ defined by
\begin{align}
    \tilde{\mu}(k_{\pm})=\mu _{\pm}, \label{WPcondition}
\end{align}
at which $B(k_0)$ changes from 1 to 0 or 0 to 1, respectively. This can occur if 
$|\mu - \mu _{\pm}|<|{2t_z}|$ and $k_\pm$ satisfy 
\begin{align}
    k_{\pm} c = \arccos \frac{\mu - \mu _{\pm}}{2t_z}. \label{WPlocation}
\end{align}
Topological phase boundaries are given by
\begin{align}
    |\mu - \mu _{\pm}|=|{2t_z}|. \label{phaseboundary}
\end{align}

\begin{figure}
	\includegraphics[width=8.6cm]{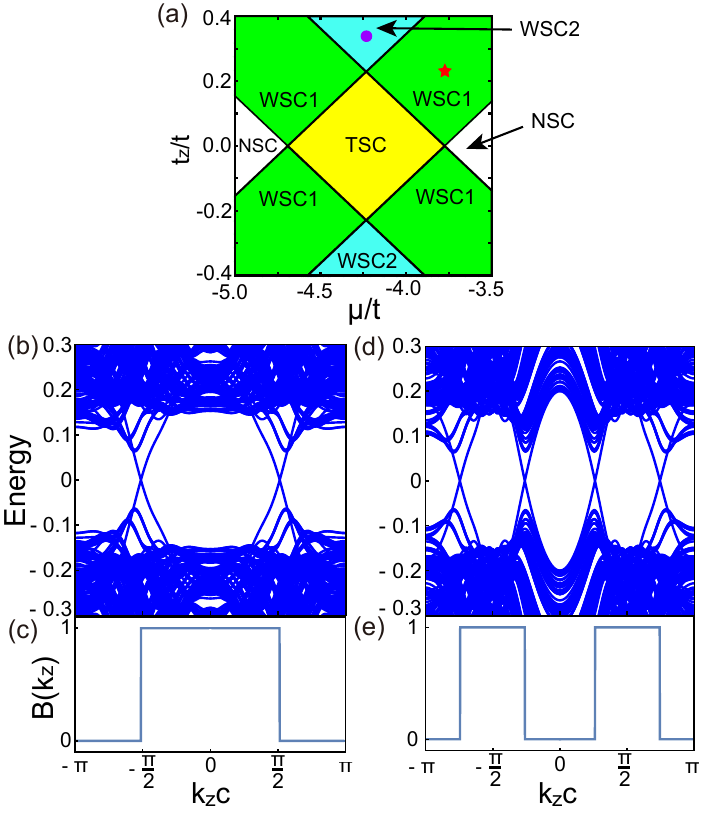}
	\caption{\label{ABcal}
    (a) Topological phase diagram for a layered superconducting Ammann-Beenker QC.
    (b) Band structure of the model in Eq.~(\ref{3DBdG}) with PBC in the $x$, $y$, and $z$ directions on layered 1393-site Ammann-Beenker approximants 
    for the parameter set marked by the red star in (a).
    (c) Bott index $B(k_z)$ as a function of $k_zc$, where $c$ is the interlayer distance, for the same system as in (b).
    (d) Same as (b) but for the parameter set marked by the purple circle in (a).
    (e) Same as (c) but for the same system as in (d).
 }
\end{figure}

Figure~\ref{ABcal}(a) shows a resulting topological phase diagram.
When $t_z=0$, each layer 
is completely isolated with no interlayer hopping and 
the topological phase diagram in Fig.~\ref{ABcal}(a) reduces to 
Fig.~\ref{AB}(b).
As in Fig.~\ref{AB}(b), the yellow region denoted as TSC in Fig.~\ref{ABcal}(a) represents a topological superconducting phase with $B(k_z)=1$, where Majorana zero-energy modes appear at surfaces, for arbitrary $k_z$ (for nonzero $t_z$).
In addition, when $t_z\neq 0$ two new topological phases emerge, 
denoted as WSC1 (green) and WSC2 (light blue) in Fig.~\ref{ABcal}(a).
The system in the WSC1 and WSC2 phase possesses, respectively, one $|k_0|$ which is either $|k_+|$ or $|k_-|$ and two $|k_0|$'s, $|k_0|=|k_+|$ and $|k_0|=|k_-|$.
Therefore, the WSC1 (WSC2) phase is a quasicrystalline Weyl superconducting phase with a pair (two pairs) of quasicrystalline Weyl nodes.
The red star (purple circle) in Fig.~\ref{ABcal}(a) indicates a representative parameter set, $\mu = \mu _+$ and $t_z=(\mu _+-\mu _-)/4$ [$\mu = (\mu _++\mu _-)/2$ and $t_z=3(\mu _+-\mu _-)/8$], which is used for studying the WSC1 (WSC2) phase below.

We first examine the bulk band structure of our 3D model Eq.~(\ref{3DBdG}) calculated with PBC in all of 
the $x$, $y$, and $z$ directions. 
The parameter set marked by the red star in Fig.~\ref{ABcal}(a) in the WSC1 phase leads to the band structure shown in Fig.~\ref{ABcal}(b), with a pair of quasicrystalline Weyl nodes at $k_0c=\pm \pi/2$.
In Fig.~\ref{ABcal}(c) $B(k_z)$ is plotted as a function of $k_z$, and we find $\chi=1$ ($-1$) for the Weyl node at $k_0 c=-\pi/2$ ($\pi/2$).
With the parameter set marked by the purple circle in Fig.~\ref{ABcal}(a) in the WSC2 phase,
the band structure exhibits two pairs of quasicrystalline Weyl nodes as can be seen in Fig.~\ref{ABcal}(d). 
The $k_z$ dependence of $B(k_z)$ in Fig.~\ref{ABcal}(e) shows that two nodes have $\chi=1$ and the other two have $\chi=-1$. 
We note that quasicrystalline Weyl nodes with opposite signs in $\chi$ 
necessarily appear in pairs, so that when the system has a Weyl node with $\chi=+1$, another Weyl node with $\chi=-1$ exists. 

\subsection{Quasiperiodic Majorana arcs}
\label{sec:results_majoranaArcs}
\begin{figure}
    \includegraphics[width=8.6cm]{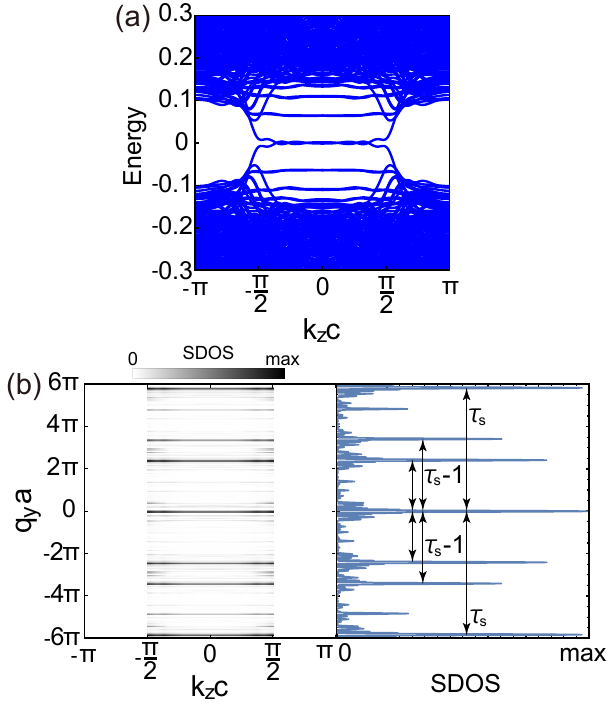}
    \caption{\label{Majorana_AB_1}
    (a) Band structure of the same system as in Fig.~\ref{ABcal}(b),(c) but with OBC in the $x$ direction, where zero-energy Majorana surface modes can be seen between the Weyl nodes at $k_zc=\pm \pi/2$.
    (b) SDOS at zero energy $\rho (q_x=0, q_y, k_z, \varepsilon =0)$ in the $q_ya$-$k_zc$ plane (left panel) and $\rho (0, q_y, k_z=0, 0)$ as a function of $q_ya$ (right panel) for the same system as in (a). Here $a$ is the link length in the 2D QC and 
    the logarithm of SDOS normalized by its maximum value is shown. 
    The ratio of $|q_ya|$ at which SDOS is the third largest (denoted as 1) to $|q_ya|$ for the second and fourth largest peaks is found to be the silver ratio $\tau_s=1+\sqrt{2}$ and $\tau_s-1$, respectively.
}
\end{figure}

\begin{figure}
    \includegraphics[width=8.6cm]{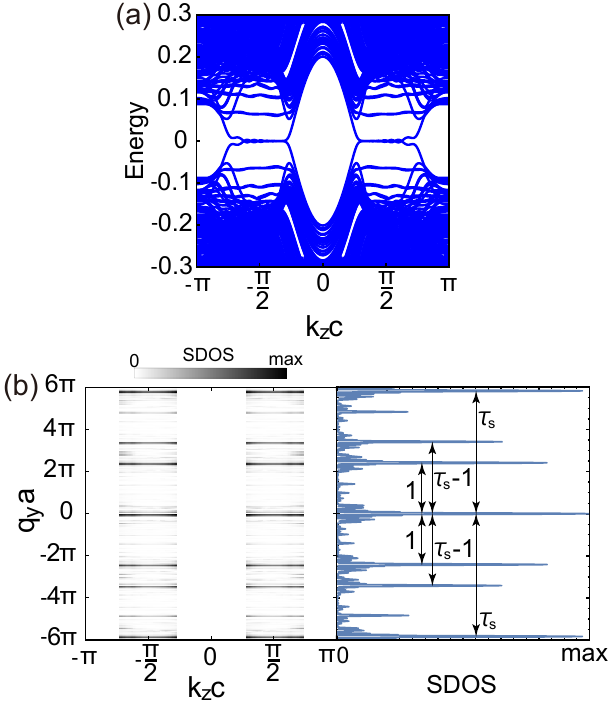}
    \caption{\label{Majorana_AB_2}
    (a) Same as Fig.~\ref{Majorana_AB_1}(a) but for the parameter set marked by the purple circle in Fig.~\ref{ABcal}(a). Two branches of Majorana zero modes can be seen between each of the two pairs of Weyl nodes.
    (b) Same as Fig.~\ref{Majorana_AB_1}(b) but for the same system as in (a) and zero-energy SDOS at $k_z c=\pi/2$ in the right panel. Similarly to the system shown in Fig.~\ref{Majorana_AB_1}, the distribution of major peaks is governed by the silver ratio.
}
\end{figure}

It can be seen in Fig.~\ref{ABcal}(c) that $B(k_z)=1$ for $-\pi/2 < k_zc < \pi/2$ between the two Weyl nodes, and similarly in Fig.~\ref{ABcal}(e) that $B(k_z)=1$ between a pair of Weyl nodes for $k_z<0$ and between another pair for $k_z>0$.
We have confirmed the existence of zero-energy Majorana surface modes for $k_z$ between a pair 
of Weyl nodes where $B(k_z)\ne 0$, just as in crystalline WSCs 
and consistently with the bulk-boundary correspondence.
We present the band structure of our system 
with OBC 
in the $x$ direction and PBC in the $y$ and $z$ directions
in Fig.~\ref{Majorana_AB_1}(a) [Fig.~\ref{Majorana_AB_2}(a)], 
for the parameter set denoted by the red star (purple circle) in Fig.~\ref{ABcal}(a). 
Figures~\ref{Majorana_AB_1}(a) and \ref{Majorana_AB_2}(a) clearly show the presence of zero-energy modes between the two Weyl nodes and each pair of Weyl nodes, respectively. These modes are Majorana surface bound states with equal-magnitude electron and hole amplitudes. 

Furthermore, we calculate the spectral density of states (SDOS) defined as
\begin{align}
    \rho (q_x, q_y, k_z, \varepsilon )
    = \sum _{m,\sigma,\tau}
    |\braket{q_x, q_y, \sigma, \tau |\psi _m(k_z)}|^2
    \delta (\varepsilon -E_m),
    \label{eq:SDOS}
\end{align}
where $q_x$ ($q_y$) is Fourier momentum in the $x$ ($y$) direction, $\varepsilon$ is energy, $\tau(=e,h)$ indicates electron or hole, $\ket{\psi _m(k_z)}$ and $E_m$ are eigenstates and eigenenergies, respectively, of the 3D Bogoliubov-de-Gennes (BdG) Hamiltonian matrix in Eq.~(\ref{3DBdG}) for a given $k_z$. 
The projection of eigenstate $\ket{\psi _m(k_z)}$, 
\begin{align}
\braket{q_x, q_y, \sigma, \tau |\psi_m(k_z)} = \frac{1}{\sqrt{N}}\sum_{r} e^{-i(q_x,q_y)\cdot\bm{R}_r}\psi_{m,\sigma,\tau}(\bm{R}_r,k_z)
\end{align}
for a given $\sigma$ and $\tau=e$ ($h$), where $N$ is the total number of vertices and $\psi_{m,\sigma,\tau}(\bm{R}_r,k_z)$ is the quasiparticle wave function at vertex coordinate $\bm{R}_r$ for a given $k_z$, is the Fourier amplitude for momentum $(q_x,q_y,k_z)$ of the electron (hole) component of the eigenstate with spin $\sigma$. 
SDOS 
has been used to investigate the effects of band structure on superconductivity in QCs \cite{Araujo19, Ghadimi21}. Experimentally, SDOS can be probed by angle-resolved photoemission spectroscopy.

The left panel in Fig.~\ref{Majorana_AB_1}(b) [\ref{Majorana_AB_2}(b)] shows SDOS at zero energy in the $q_ya$-$k_zc$ plane, $\rho (q_x=0, q_y, k_z, \varepsilon =0)$, where $a$ is the link length in the 2D QC, for the parameter set marked by the red star (purple circle) in Fig.~\ref{ABcal}(a). 
The zero-energy SDOS is peaked at a series of $q_ya$ values, forming lines between the two 
(each pair of) Weyl nodes. 
By analogy with ``Fermi-Bragg'' arcs in quasicrystalline Weyl semimetals \cite{Grossi23}, we call these lines Majorana arcs, which are distributed densely and quasiperiodically as a function of $q_ya$.
Quasiperiodic Majorana arcs are analogous to Majorana arcs in crystals repeating in momentum space, where the repetition corresponds to the periodicity of Bragg peaks (see Appendix~\ref{square}). Since Bragg peaks in a QC exhibit quasiperiodicity, its Majorana arcs, e.g., at $q_y=0$ are quasiperiodically repeated in momentum space. We note that the arcs are all lines owing to the simplicity of our model \cite{Grossi23}. This is illustrated in the right panels of Figs.~\ref{Majorana_AB_1}(b) and \ref{Majorana_AB_2}(b), where
$\rho (0, q_y, k_z=0, 0)$ and $\rho (0, q_y, k_zc=\pi/2, 0)$, respectively, are plotted as a function of $q_ya$ and can be seen to be the largest at and symmetric about $q_y=0$. The ratio of the smallest $|q_ya|\ne 0$ at which SDOS is substantially large (the third largest in the range shown), denoted as 1, to $|q_ya|$ for the second and fourth largest peaks in SDOS is found to be the silver ratio, $\tau_s=1+\sqrt{2}$, and $\tau_s-1$, respectively.
The silver ratio $\tau_s$ is associated with the structure of Ammann-Beenker QCs \cite{Janot_1992, Koga_2020}.
We have checked for major peak (${\rm SDOS} > {\rm max}/2$) positions in a range much larger than shown in Figs.~\ref{Majorana_AB_1}(b) and \ref{Majorana_AB_2}(b) ($|q_ya|\le 50\pi$) that the ratio of a given $|q_ya|$ at which SDOS is peaked to another $|q_ya|$ where SDOS has another peak is given by $\alpha \tau_s +\beta$, where $\alpha$ and $\beta$ are rational numbers. 
Since $\alpha \tau_s +\beta$ is an irrational number, the distribution of Majorana arcs is quasiperiodic in momentum in the direction of surfaces within each quasicrystalline layer. This makes sense, as we have also checked that the positions of major peaks in SDOS in the $q_y$ axis match those of major Bragg peaks of Ammann-Beenker QCs. Like the Bragg peaks, with higher resolution, an infinite number of Majorana arcs will fill in the entire $q_y$ space.

Weyl superconductivity can also occur in periodically layered Penrose QCs. It is shown in Appendix~\ref{Penrose} that Majorana arcs appear between a pair of Weyl nodes and their quasiperiodic distribution in momentum along each surface of quasicrystalline layers 
is characterized by the golden ratio, $\tau_g=(1+\sqrt{5})/2$, which is inherent in the structure of Penrose QCs.
The quasiperiodic distribution of Majorana arcs in quasicrystalline WSCs is in striking contrast to the periodic distribution of Majorana arcs in momentum space of crystalline WSCs.
In Appendix~\ref{square}, we demonstrate periodic Majorana arcs in 
a layered square-lattice Weyl superconductor. 

\section{Summary}
\label{sec:summary}

In summary, we have demonstrated the occurrence of quasicrystalline Weyl superconductivity.
We have generalized the concept of Weyl superconductivity in 3D crystals to periodically stacked 2D QCs, by identifying quasicrystalline Weyl nodes in one-dimensional momentum space corresponding to the stacking direction, where the Bott index changes its value. 
Between two quasicrystalline Weyl nodes in momentum space, where the Bott index is nonzero, Majorana zero modes appear on surfaces.
By calculating SDOS, 
we have found that these Majorana arcs appear densely and quasiperiodically as a function of momentum
in the direction of surfaces within each 2D QC, whose quasiperiodicity is governed by the silver and golden ratio, respectively, for Ammann-Beenker and Penrose QCs. 

One of the candidate materials to be a quasicrystalline WSC is a van der Waals-layered Ta$_{1.6}$Te QC \cite{Conrad_1998} whose structure consists of layered 2D dodecagonal QCs.
Recently, Tokumoto {\it et al.} \cite{Tokumoto23} have observed bulk superconductivity in Ta$_{1.6}$Te QCs.
This finding expands the possibility of realization of quasicrystalline Weyl superconductivity.

\begin{acknowledgments}
    We thank Yuki Tokumoto for useful discussions.
    This work was supported by JSPS KAKENHI (Grant No. 23K13033), JST SPRING (Grant No. JPMJSP2151), and the Natural Sciences and Engineering Research Council of Canada.
\end{acknowledgments}

\appendix
\section{Bott index} \label{Bott}
We review the Bott index $B$ \cite{Loring10} which is an integer topological invariant. 
Whether the system is periodic or not, 
the Bott index $B$ topologically classifies phases of a 2D system whose Hamiltonian belongs to class $D$, including time-reversal-breaking superconductors \cite{Altland_1997, Schnyder_2008, Loring10, Chiu_2016, Ghadimi21}. In a superconducting state with $B\neq 0$, Majorana zero-energy modes can appear at surfaces or topological defects. 

The Bott index is given by
\begin{align}
    B = \frac{1}{2\pi}\mathrm{Im}\mathrm{Tr}[\log (U_YU_XU_Y^{\dagger}U_X^{\dagger})],
\end{align}
where $U_X$ ($U_Y$) is a projected position operator in the $x$ ($y$) direction. Using an occupation projector that is defined as
\begin{align}
    P=\sum _{E_{m}<0}\ket{\psi_m}\bra{\psi_m},
\end{align}
the projected position operators are given by
\begin{align}
    U_X=Pe^{2\pi i X}P+(I-P), \hspace{3mm}
    U_Y=Pe^{2\pi i Y}P+(I-P),
\end{align}
where $\ket{\psi _m}$ is an eigenstate of the Bogoliubov-de-Gennes Hamiltonian for a given system with a negative eigenvalue $E_{m}$, $X$ ($Y$) is a rescaled position operator defined in the interval $[0, 1)$ in the $x$ ($y$) direction, and $I$ is an identity operator. 
When the spectral gap of the system closes as a system parameter is varied, 
the value of $B$ can change.
In the thermodynamic limit, any 2D system with OBC has $B=0$.
Therefore, we use approximants of Ammann-Beenker and Penrose QCs with PBC in both directions to calculate $B$.

\section{Quasicrystalline Weyl superconductivity in layered Penrose QCs} \label{Penrose}
\begin{figure}
	\includegraphics[width=8.6cm]{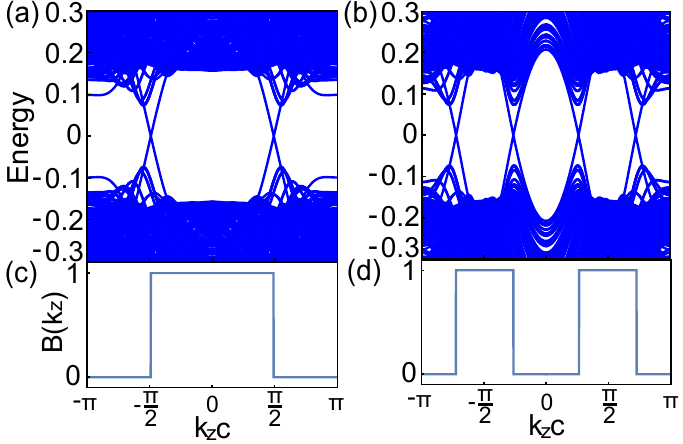}
	\caption{\label{Pencal_1}
    (a) Band structure of the model in Eq.~(\ref{3DBdG}) for layered 3571-site Penrose approximants with PBC in the $x$, $y$, and $z$ directions, for 
    $\mu = \mu _+$ and $t_z=(\mu _+-\mu _-)/4$.
    (b) Same as (a) but for $\mu = (\mu _++\mu _-)/2$ and $t_z=3(\mu _+-\mu _-)/8$.
    (c) The Bott index $B(k_z)$ as a function of $k_zc$ for the same system as in (a).
    (d) Same as (c) but for the same system as in (b).
 }
\end{figure}

We consider 
periodically stacked, superconducting Penrose QCs 
whose Hamiltonian is given by Eq.~(\ref{3DBdG}).
For Penrose QCs, by using $\mu _{+}/t \simeq -3.78$ and $\mu _{-}/t \simeq -4.69$ \cite{Ghadimi21},
we obtain a topological phase diagram similar to Fig.~\ref{ABcal}(a). 
We choose two parameter sets given by $\mu = \mu _+$ and $t_z=(\mu _+-\mu _-)/4$, and $\mu = (\mu _++\mu _-)/2$ and $t_z=3(\mu _+-\mu _-)/8$,
to calculate the band structure which is shown in Figs.~\ref{Pencal_1}(a) and \ref{Pencal_1}(b), respectively.
Figure~\ref{Pencal_1}(a) [\ref{Pencal_1}(b)] shows the band structure of the model with PBC in the $x$, $y$, and $z$ directions, 
where we find a pair (two pairs) of quasicrystalline Weyl nodes. 
It can be seen from the Bott index $B(k_z)$ shown in Figs.~\ref{Pencal_1}(c) and \ref{Pencal_1}(d) that $\chi=+1$ or $-1$ at each of these Weyl nodes. 

\begin{figure}
    \includegraphics[width=8.6cm]{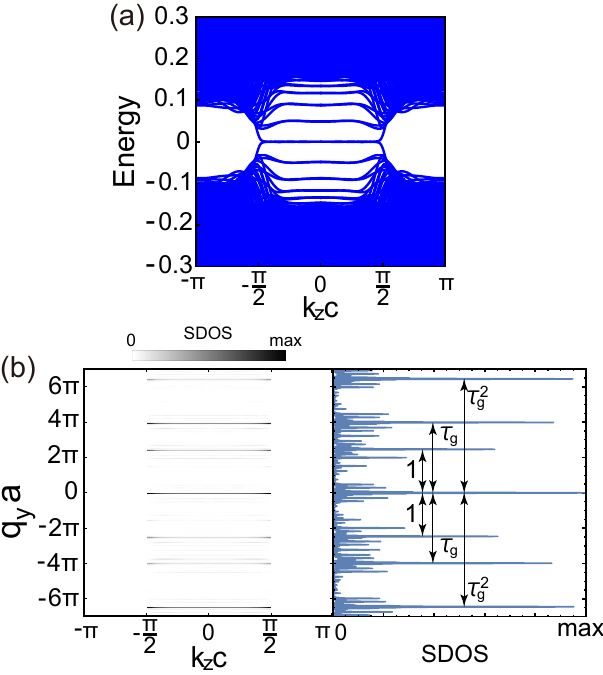}
	\caption{\label{Pencal_2}
    (a) Band structure of the same system as in Fig.~\ref{Pencal_1}(a),~\ref{Pencal_1}(c) but with OBC in the $x$ direction, showing Majorana surface states between the two Weyl nodes at $k_zc=\pm \pi/2$.
    (b) Logarithm of SDOS at zero energy $\rho (q_x=0, q_y, k_z, \varepsilon =0)$ in the $q_ya$-$k_zc$ plane (left panel) and $\rho (0, q_y, k_z=0, 0)$ as a function of $q_ya$ (right panel), where $a$ is the link length in the 2D QC, for the same system as in (a). 
    The ratio of $|q_ya|$ at which SDOS is the fourth largest (denoted as 1) to $|q_ya|$ for the third and second largest peaks is found to be the golden ratio $\tau_g=(1+\sqrt{5})/2$ and $\tau_g^2$, respectively.
 }
\end{figure}

\begin{figure}
    \includegraphics[width=8.6cm]{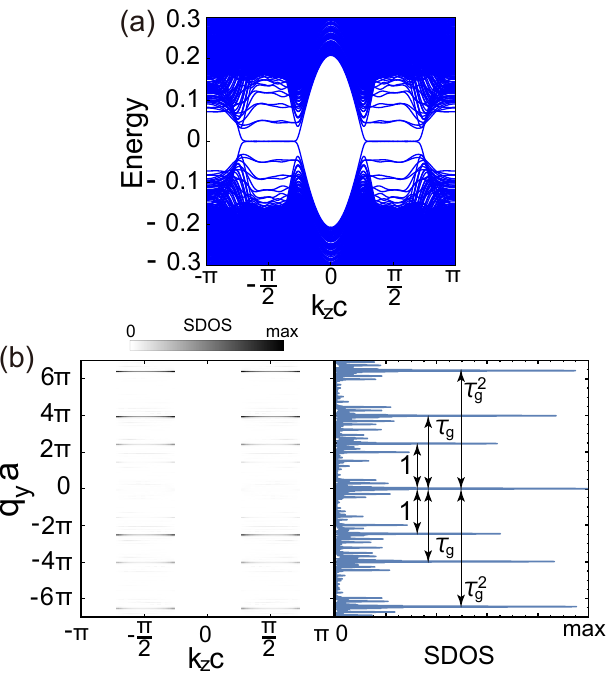} 
	\caption{\label{Pencal_3}
    (a) Same as Fig.~\ref{Pencal_2}(a) but for 
    $\mu = (\mu _++\mu _-)/2$ and $t_z=3(\mu _+-\mu _-)/8$, where Majorana zero modes appear between each pair of Weyl nodes. 
    (b) Same as Fig.~\ref{Pencal_2}(b) but for the same system as in (a) and zero-energy SDOS at $k_z c=\pi/2$ in the right panel. Similarly to the system shown in Fig.~\ref{Pencal_2}, the distribution of major peaks is governed by the golden ratio.
    }
\end{figure}

Figure~\ref{Pencal_2}(a) [\ref{Pencal_3}(a)] shows the band structure of the model with OBC in the $x$ direction and PBC in the $y$ and $z$ directions for the first (second) parameter set above.
Zero-energy modes appear at surfaces between two quasicrystalline Weyl nodes in $k_z$ space, where the Bott index is nonzero.
We have calculated SDOS in Eq.~(\ref{eq:SDOS}) and present $\rho (q_x=0, q_y, k_z, \varepsilon =0)$ in the left panel of Fig.~\ref{Pencal_2}(b) [\ref{Pencal_3}(b)], where 
Majorana arcs can be seen between the two (each pair of) Weyl nodes 
at densely and quasiperiodically distributed positions in $q_ya$.
This is illustrated further in terms of $\rho (0, q_y, k_z=0, 0)$ and $\rho (0, q_y, k_zc=\pi/2, 0)$ shown in the right panels of Figs.~\ref{Pencal_2}(b) and \ref{Pencal_3}(b), respectively. As in Ammann-Beenker QCs, SDOS is the largest at and symmetric about $q_y=0$.
The ratio of the smallest $|q_ya|\ne 0$ at which SDOS has a major peak (the fourth highest in the range shown) to $|q_ya|$ for the third and second highest peaks is found to be, respectively, $1:\tau_g$ and $1:{\tau_g}^2$, where $\tau_g=(1+\sqrt{5})/2$ is the golden ratio.
Geometric and physical properties of Penrose QCs are governed by 
$\tau_g$ \cite{Janot_1992, Gambaudo_2014, Koga_2017}.
We have confirmed for $|q_ya|\le 50\pi$ that the ratio of $|q_ya|$ for a major peak (${\rm SDOS} > {\rm max}/2$) in SDOS to $|q_ya|$ for another major peak can be expressed as $1:\alpha \tau_g + \beta$, where $\alpha$ and $\beta$ are rational numbers, and that those peak positions coincide with $q_ya$ values of major Bragg peaks of Penrose QCs.
Hence, similarly to Ammann-Beenker QCs, layered Penrose QCs exhibit quasicrystalline Weyl superconductivity, with Majorana arcs between a pair of quasicrystalline Weyl nodes positioned quasiperiodically in momentum in the direction of surfaces within each layer.


\section{Comparison with a crystalline WSC} \label{square}
\begin{figure}
	\includegraphics[width=8.6cm]{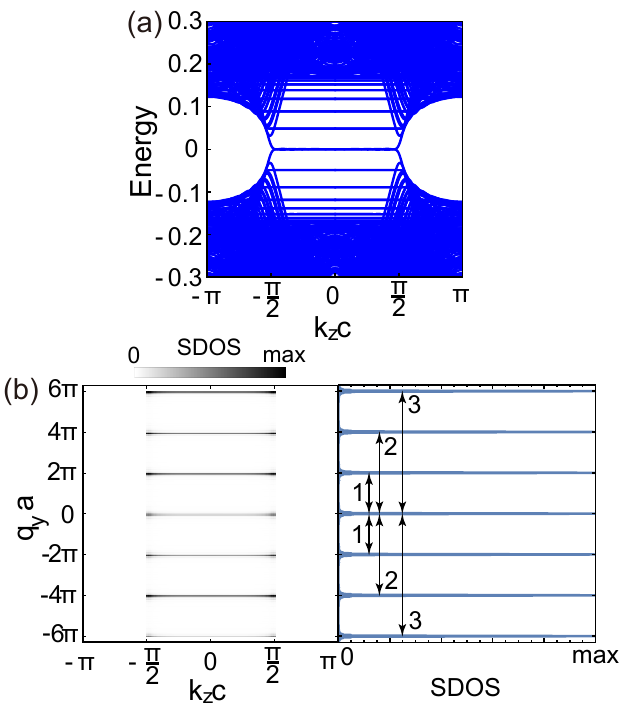}
    \caption{\label{Majorana_Sq_1} 
    (a) Band structure of a layered square-lattice Weyl superconductor with OBC in the $x$ direction and PBC in the $y$ and $z$ directions, for 
    $\mu = \mu _+$ and $t_z=(\mu _+-\mu _-)/4$.
    (b) Logarithm of SDOS at zero energy $\rho (q_x=0, q_y, k_z, \varepsilon =0)$ in the $q_ya$-$k_zc$ plane (left panel) and $\rho (0, q_y, k_z=0, 0)$ as a function of $q_ya$ (right panel)
    for the system as in (a), where $a$ is the lattice constant of the square lattice. 
    Each layer has $50\times 50$ lattice sites.
    }
\end{figure}

We now study a layered square-lattice superconductor whose Hamiltonian is given by Eq.~(\ref{3DBdG}).
For a square lattice, topological phase boundaries are given analytically by $\mu_{\pm}=-4t\pm \sqrt{h_z^2-\Delta_0^2}$ in the low-filling limit \cite{Sato09,Sato10}, which yield 
a topological phase diagram similar to Fig.~\ref{ABcal}(a).
For the parameter set $\mu = \mu _+$ and $t_z=(\mu _+-\mu _-)/4$,
the band structure shows one pair of Weyl nodes with $\chi=\pm 1$ (at $k_zc=\mp \pi/2$), which is chirality of each Weyl node in this case.
The band structure of the model with OBC in the $x$ direction and PBC in the $y$ and $z$ directions is presented in Fig.~\ref{Majorana_Sq_1}(a), where zero-energy surface states can be seen.
The corresponding SDOS $\rho (q_x=0, q_y, k_z, \varepsilon =0)$ 
is shown in the left panel of Fig.~\ref{Majorana_Sq_1}(b), where Majorana arcs appear between the two Weyl nodes where the Bott index is unity.
As illustrated in the right panel of Fig.~\ref{Majorana_Sq_1}(b) in terms of $\rho (0, q_y, k_z=0, 0)$, the Majorana arcs are equally spaced in $q_y$ and positioned 
at $q_y a=0, \pm 2\pi, \pm 4\pi, \pm 6\pi \ldots$.
Therefore, any distance from a given Majorana arc to another Majorana arc is a natural number multiple of $2\pi$ and their ratio is a 
rational number, as indicated in the right panel of Fig.~\ref{Majorana_Sq_1}(b).
This contrasts with the quasiperiodic distribution of Majorana arcs in quasicrystalline WSCs, where an irrational number characterizes those ratios.

\bibliography{QWSC}

\begin{thebibliography}{66}%
\makeatletter
\providecommand \@ifxundefined [1]{%
 \@ifx{#1\undefined}
}%
\providecommand \@ifnum [1]{%
 \ifnum #1\expandafter \@firstoftwo
 \else \expandafter \@secondoftwo
 \fi
}%
\providecommand \@ifx [1]{%
 \ifx #1\expandafter \@firstoftwo
 \else \expandafter \@secondoftwo
 \fi
}%
\providecommand \natexlab [1]{#1}%
\providecommand \enquote  [1]{``#1''}%
\providecommand \bibnamefont  [1]{#1}%
\providecommand \bibfnamefont [1]{#1}%
\providecommand \citenamefont [1]{#1}%
\providecommand \href@noop [0]{\@secondoftwo}%
\providecommand \href [0]{\begingroup \@sanitize@url \@href}%
\providecommand \@href[1]{\@@startlink{#1}\@@href}%
\providecommand \@@href[1]{\endgroup#1\@@endlink}%
\providecommand \@sanitize@url [0]{\catcode `\\12\catcode `\$12\catcode
  `\&12\catcode `\#12\catcode `\^12\catcode `\_12\catcode `\%12\relax}%
\providecommand \@@startlink[1]{}%
\providecommand \@@endlink[0]{}%
\providecommand \url  [0]{\begingroup\@sanitize@url \@url }%
\providecommand \@url [1]{\endgroup\@href {#1}{\urlprefix }}%
\providecommand \urlprefix  [0]{URL }%
\providecommand \Eprint [0]{\href }%
\providecommand \doibase [0]{https://doi.org/}%
\providecommand \selectlanguage [0]{\@gobble}%
\providecommand \bibinfo  [0]{\@secondoftwo}%
\providecommand \bibfield  [0]{\@secondoftwo}%
\providecommand \translation [1]{[#1]}%
\providecommand \BibitemOpen [0]{}%
\providecommand \bibitemStop [0]{}%
\providecommand \bibitemNoStop [0]{.\EOS\space}%
\providecommand \EOS [0]{\spacefactor3000\relax}%
\providecommand \BibitemShut  [1]{\csname bibitem#1\endcsname}%
\let\auto@bib@innerbib\@empty
\bibitem [{\citenamefont {Shechtman}\ \emph {et~al.}(1984)\citenamefont
  {Shechtman}, \citenamefont {Blech}, \citenamefont {Gratias},\ and\
  \citenamefont {Cahn}}]{Shechtman_1984}%
  \BibitemOpen
  \bibfield  {author} {\bibinfo {author} {\bibfnamefont {D.}~\bibnamefont
  {Shechtman}}, \bibinfo {author} {\bibfnamefont {I.}~\bibnamefont {Blech}},
  \bibinfo {author} {\bibfnamefont {D.}~\bibnamefont {Gratias}},\ and\ \bibinfo
  {author} {\bibfnamefont {J.~W.}\ \bibnamefont {Cahn}},\ }\bibfield  {title}
  {\bibinfo {title} {Metallic phase with long-range orientational order and no
  translational symmetry},\ }\href
  {https://doi.org/10.1103/PhysRevLett.53.1951} {\bibfield  {journal} {\bibinfo
   {journal} {Phys. Rev. Lett.}\ }\textbf {\bibinfo {volume} {53}},\ \bibinfo
  {pages} {1951} (\bibinfo {year} {1984})}\BibitemShut {NoStop}%
\bibitem [{\citenamefont {Tamura}\ \emph {et~al.}(2021)\citenamefont {Tamura},
  \citenamefont {Ishikawa}, \citenamefont {Suzuki}, \citenamefont {Kotajima},
  \citenamefont {Tanaka}, \citenamefont {Seki}, \citenamefont {Shibata},
  \citenamefont {Yamada}, \citenamefont {Fujii}, \citenamefont {Wang},
  \citenamefont {Avdeev}, \citenamefont {Nawa}, \citenamefont {Okuyama},\ and\
  \citenamefont {Sato}}]{Tamura_2021}%
  \BibitemOpen
  \bibfield  {author} {\bibinfo {author} {\bibfnamefont {R.}~\bibnamefont
  {Tamura}}, \bibinfo {author} {\bibfnamefont {A.}~\bibnamefont {Ishikawa}},
  \bibinfo {author} {\bibfnamefont {S.}~\bibnamefont {Suzuki}}, \bibinfo
  {author} {\bibfnamefont {T.}~\bibnamefont {Kotajima}}, \bibinfo {author}
  {\bibfnamefont {Y.}~\bibnamefont {Tanaka}}, \bibinfo {author} {\bibfnamefont
  {T.}~\bibnamefont {Seki}}, \bibinfo {author} {\bibfnamefont {N.}~\bibnamefont
  {Shibata}}, \bibinfo {author} {\bibfnamefont {T.}~\bibnamefont {Yamada}},
  \bibinfo {author} {\bibfnamefont {T.}~\bibnamefont {Fujii}}, \bibinfo
  {author} {\bibfnamefont {C.-W.}\ \bibnamefont {Wang}}, \bibinfo {author}
  {\bibfnamefont {M.}~\bibnamefont {Avdeev}}, \bibinfo {author} {\bibfnamefont
  {K.}~\bibnamefont {Nawa}}, \bibinfo {author} {\bibfnamefont {D.}~\bibnamefont
  {Okuyama}},\ and\ \bibinfo {author} {\bibfnamefont {T.~J.}\ \bibnamefont
  {Sato}},\ }\bibfield  {title} {\bibinfo {title} {Experimental observation of
  long-range magnetic order in icosahedral quasicrystals},\ }\href
  {https://doi.org/10.1021/jacs.1c09954} {\bibfield  {journal} {\bibinfo
  {journal} {J. Am. Chem. Soc.}\ }\textbf {\bibinfo {volume} {143}},\ \bibinfo
  {pages} {19938} (\bibinfo {year} {2021})}\BibitemShut {NoStop}%
\bibitem [{\citenamefont {Kamiya}\ \emph {et~al.}(2018)\citenamefont {Kamiya},
  \citenamefont {Takeuchi}, \citenamefont {Kabeya}, \citenamefont {Wada},
  \citenamefont {Ishimasa}, \citenamefont {Ochiai}, \citenamefont {Deguchi},
  \citenamefont {Imura},\ and\ \citenamefont {Sato}}]{Kamiya_2018}%
  \BibitemOpen
  \bibfield  {author} {\bibinfo {author} {\bibfnamefont {K.}~\bibnamefont
  {Kamiya}}, \bibinfo {author} {\bibfnamefont {T.}~\bibnamefont {Takeuchi}},
  \bibinfo {author} {\bibfnamefont {N.}~\bibnamefont {Kabeya}}, \bibinfo
  {author} {\bibfnamefont {N.}~\bibnamefont {Wada}}, \bibinfo {author}
  {\bibfnamefont {T.}~\bibnamefont {Ishimasa}}, \bibinfo {author}
  {\bibfnamefont {A.}~\bibnamefont {Ochiai}}, \bibinfo {author} {\bibfnamefont
  {K.}~\bibnamefont {Deguchi}}, \bibinfo {author} {\bibfnamefont
  {K.}~\bibnamefont {Imura}},\ and\ \bibinfo {author} {\bibfnamefont {N.~K.}\
  \bibnamefont {Sato}},\ }\bibfield  {title} {\bibinfo {title} {Discovery of
  superconductivity in quasicrystal},\ }\href
  {https://doi.org/10.1038/s41467-017-02667-x} {\bibfield  {journal} {\bibinfo
  {journal} {Nat. Commun.}\ }\textbf {\bibinfo {volume} {9}},\ \bibinfo {pages}
  {154} (\bibinfo {year} {2018})}\BibitemShut {NoStop}%
\bibitem [{\citenamefont {Tokumoto}\ \emph {et~al.}(2024)\citenamefont
  {Tokumoto}, \citenamefont {Hamano}, \citenamefont {Nakagawa}, \citenamefont
  {Kamimura}, \citenamefont {Suzuki}, \citenamefont {Tamura},\ and\
  \citenamefont {Edagawa}}]{Tokumoto23}%
  \BibitemOpen
  \bibfield  {author} {\bibinfo {author} {\bibfnamefont {Y.}~\bibnamefont
  {Tokumoto}}, \bibinfo {author} {\bibfnamefont {K.}~\bibnamefont {Hamano}},
  \bibinfo {author} {\bibfnamefont {S.}~\bibnamefont {Nakagawa}}, \bibinfo
  {author} {\bibfnamefont {Y.}~\bibnamefont {Kamimura}}, \bibinfo {author}
  {\bibfnamefont {S.}~\bibnamefont {Suzuki}}, \bibinfo {author} {\bibfnamefont
  {R.}~\bibnamefont {Tamura}},\ and\ \bibinfo {author} {\bibfnamefont
  {K.}~\bibnamefont {Edagawa}},\ }\bibfield  {title} {\bibinfo {title}
  {Superconductivity in a van der {W}aals layered quasicrystal},\ }\href
  {https://www.nature.com/articles/s41467-024-45952-2} {\bibfield  {journal}
  {\bibinfo  {journal} {Nat. Commun.}\ }\textbf {\bibinfo {volume} {15}},\
  \bibinfo {pages} {1529} (\bibinfo {year} {2024})}\BibitemShut {NoStop}%
\bibitem [{\citenamefont {Sakai}\ \emph {et~al.}(2017)\citenamefont {Sakai},
  \citenamefont {Takemori}, \citenamefont {Koga},\ and\ \citenamefont
  {Arita}}]{Sakai_2017}%
  \BibitemOpen
  \bibfield  {author} {\bibinfo {author} {\bibfnamefont {S.}~\bibnamefont
  {Sakai}}, \bibinfo {author} {\bibfnamefont {N.}~\bibnamefont {Takemori}},
  \bibinfo {author} {\bibfnamefont {A.}~\bibnamefont {Koga}},\ and\ \bibinfo
  {author} {\bibfnamefont {R.}~\bibnamefont {Arita}},\ }\bibfield  {title}
  {\bibinfo {title} {Superconductivity on a quasiperiodic lattice:
  Extended-to-localized crossover of {C}ooper pairs},\ }\href
  {https://doi.org/10.1103/PhysRevB.95.024509} {\bibfield  {journal} {\bibinfo
  {journal} {Phys. Rev. B}\ }\textbf {\bibinfo {volume} {95}},\ \bibinfo
  {pages} {024509} (\bibinfo {year} {2017})}\BibitemShut {NoStop}%
\bibitem [{\citenamefont {Sakai}\ and\ \citenamefont
  {Arita}(2019)}]{Sakai_2019}%
  \BibitemOpen
  \bibfield  {author} {\bibinfo {author} {\bibfnamefont {S.}~\bibnamefont
  {Sakai}}\ and\ \bibinfo {author} {\bibfnamefont {R.}~\bibnamefont {Arita}},\
  }\bibfield  {title} {\bibinfo {title} {Exotic pairing state in
  quasicrystalline superconductors under a magnetic field},\ }\href
  {https://doi.org/10.1103/PhysRevResearch.1.022002} {\bibfield  {journal}
  {\bibinfo  {journal} {Phys. Rev. Res.}\ }\textbf {\bibinfo {volume} {1}},\
  \bibinfo {pages} {022002(R)} (\bibinfo {year} {2019})}\BibitemShut {NoStop}%
\bibitem [{\citenamefont {Takemori}\ \emph {et~al.}(2020)\citenamefont
  {Takemori}, \citenamefont {Arita},\ and\ \citenamefont
  {Sakai}}]{Takemori_2020}%
  \BibitemOpen
  \bibfield  {author} {\bibinfo {author} {\bibfnamefont {N.}~\bibnamefont
  {Takemori}}, \bibinfo {author} {\bibfnamefont {R.}~\bibnamefont {Arita}},\
  and\ \bibinfo {author} {\bibfnamefont {S.}~\bibnamefont {Sakai}},\ }\bibfield
   {title} {\bibinfo {title} {Physical properties of weak-coupling
  quasiperiodic superconductors},\ }\href
  {https://doi.org/10.1103/PhysRevB.102.115108} {\bibfield  {journal} {\bibinfo
   {journal} {Phys. Rev. B}\ }\textbf {\bibinfo {volume} {102}},\ \bibinfo
  {pages} {115108} (\bibinfo {year} {2020})}\BibitemShut {NoStop}%
\bibitem [{\citenamefont {Fukushima}\ \emph
  {et~al.}(2023{\natexlab{a}})\citenamefont {Fukushima}, \citenamefont
  {Takemori}, \citenamefont {Sakai}, \citenamefont {Ichioka},\ and\
  \citenamefont {Jagannathan}}]{Fukushima_2023}%
  \BibitemOpen
  \bibfield  {author} {\bibinfo {author} {\bibfnamefont {T.}~\bibnamefont
  {Fukushima}}, \bibinfo {author} {\bibfnamefont {N.}~\bibnamefont {Takemori}},
  \bibinfo {author} {\bibfnamefont {S.}~\bibnamefont {Sakai}}, \bibinfo
  {author} {\bibfnamefont {M.}~\bibnamefont {Ichioka}},\ and\ \bibinfo {author}
  {\bibfnamefont {A.}~\bibnamefont {Jagannathan}},\ }\bibfield  {title}
  {\bibinfo {title} {Supercurrent distribution on {A}mmann-{B}eenker
  structure},\ }\href {https://doi.org/10.1088/1742-6596/2461/1/012014}
  {\bibfield  {journal} {\bibinfo  {journal} {J. Phys.: Conf. Ser.}\ }\textbf
  {\bibinfo {volume} {2461}},\ \bibinfo {pages} {012014} (\bibinfo {year}
  {2023}{\natexlab{a}})}\BibitemShut {NoStop}%
\bibitem [{\citenamefont {Fukushima}\ \emph
  {et~al.}(2023{\natexlab{b}})\citenamefont {Fukushima}, \citenamefont
  {Takemori}, \citenamefont {Sakai}, \citenamefont {Ichioka},\ and\
  \citenamefont {Jagannathan}}]{Fukushima_2023_2}%
  \BibitemOpen
  \bibfield  {author} {\bibinfo {author} {\bibfnamefont {T.}~\bibnamefont
  {Fukushima}}, \bibinfo {author} {\bibfnamefont {N.}~\bibnamefont {Takemori}},
  \bibinfo {author} {\bibfnamefont {S.}~\bibnamefont {Sakai}}, \bibinfo
  {author} {\bibfnamefont {M.}~\bibnamefont {Ichioka}},\ and\ \bibinfo {author}
  {\bibfnamefont {A.}~\bibnamefont {Jagannathan}},\ }\bibfield  {title}
  {\bibinfo {title} {Supercurrent distribution in real-space and anomalous
  paramagnetic response in a superconducting quasicrystal},\ }\href
  {https://doi.org/10.1103/PhysRevResearch.5.043164} {\bibfield  {journal}
  {\bibinfo  {journal} {Phys. Rev. Res.}\ }\textbf {\bibinfo {volume} {5}},\
  \bibinfo {pages} {043164} (\bibinfo {year} {2023}{\natexlab{b}})}\BibitemShut
  {NoStop}%
\bibitem [{\citenamefont {Fan}\ and\ \citenamefont {Huang}(2021)}]{Fan21}%
  \BibitemOpen
  \bibfield  {author} {\bibinfo {author} {\bibfnamefont {J.}~\bibnamefont
  {Fan}}\ and\ \bibinfo {author} {\bibfnamefont {H.}~\bibnamefont {Huang}},\
  }\bibfield  {title} {\bibinfo {title} {Topological states in quasicrystals},\
  }\href {https://doi.org/10.1007/s11467-021-1100-y} {\bibfield  {journal}
  {\bibinfo  {journal} {Front. Phys.}\ }\textbf {\bibinfo {volume} {17}},\
  \bibinfo {pages} {13203} (\bibinfo {year} {2021})}\BibitemShut {NoStop}%
\bibitem [{\citenamefont {Tran}\ \emph {et~al.}(2015)\citenamefont {Tran},
  \citenamefont {Dauphin}, \citenamefont {Goldman},\ and\ \citenamefont
  {Gaspard}}]{Tran15}%
  \BibitemOpen
  \bibfield  {author} {\bibinfo {author} {\bibfnamefont {D.-T.}\ \bibnamefont
  {Tran}}, \bibinfo {author} {\bibfnamefont {A.}~\bibnamefont {Dauphin}},
  \bibinfo {author} {\bibfnamefont {N.}~\bibnamefont {Goldman}},\ and\ \bibinfo
  {author} {\bibfnamefont {P.}~\bibnamefont {Gaspard}},\ }\bibfield  {title}
  {\bibinfo {title} {Topological {H}ofstadter insulators in a two-dimensional
  quasicrystal},\ }\href {https://doi.org/10.1103/PhysRevB.91.085125}
  {\bibfield  {journal} {\bibinfo  {journal} {Phys. Rev. B}\ }\textbf {\bibinfo
  {volume} {91}},\ \bibinfo {pages} {085125} (\bibinfo {year}
  {2015})}\BibitemShut {NoStop}%
\bibitem [{\citenamefont {Bandres}\ \emph {et~al.}(2016)\citenamefont
  {Bandres}, \citenamefont {Rechtsman},\ and\ \citenamefont
  {Segev}}]{Bandres16}%
  \BibitemOpen
  \bibfield  {author} {\bibinfo {author} {\bibfnamefont {M.~A.}\ \bibnamefont
  {Bandres}}, \bibinfo {author} {\bibfnamefont {M.~C.}\ \bibnamefont
  {Rechtsman}},\ and\ \bibinfo {author} {\bibfnamefont {M.}~\bibnamefont
  {Segev}},\ }\bibfield  {title} {\bibinfo {title} {Topological photonic
  quasicrystals: Fractal topological spectrum and protected transport},\ }\href
  {https://doi.org/10.1103/PhysRevX.6.011016} {\bibfield  {journal} {\bibinfo
  {journal} {Phys. Rev. X}\ }\textbf {\bibinfo {volume} {6}},\ \bibinfo {pages}
  {011016} (\bibinfo {year} {2016})}\BibitemShut {NoStop}%
\bibitem [{\citenamefont {Huang}\ and\ \citenamefont
  {Liu}(2018{\natexlab{a}})}]{Huang18PRL}%
  \BibitemOpen
  \bibfield  {author} {\bibinfo {author} {\bibfnamefont {H.}~\bibnamefont
  {Huang}}\ and\ \bibinfo {author} {\bibfnamefont {F.}~\bibnamefont {Liu}},\
  }\bibfield  {title} {\bibinfo {title} {Theory of spin {B}ott index for
  quantum spin {H}all states in nonperiodic systems},\ }\href
  {https://doi.org/10.1103/PhysRevB.98.125130} {\bibfield  {journal} {\bibinfo
  {journal} {Phys. Rev. B}\ }\textbf {\bibinfo {volume} {98}},\ \bibinfo
  {pages} {125130} (\bibinfo {year} {2018}{\natexlab{a}})}\BibitemShut
  {NoStop}%
\bibitem [{\citenamefont {Huang}\ and\ \citenamefont
  {Liu}(2018{\natexlab{b}})}]{Huang18PRB}%
  \BibitemOpen
  \bibfield  {author} {\bibinfo {author} {\bibfnamefont {H.}~\bibnamefont
  {Huang}}\ and\ \bibinfo {author} {\bibfnamefont {F.}~\bibnamefont {Liu}},\
  }\bibfield  {title} {\bibinfo {title} {Quantum spin {H}all effect and spin
  {B}ott index in a quasicrystal lattice},\ }\href
  {https://doi.org/10.1103/PhysRevLett.121.126401} {\bibfield  {journal}
  {\bibinfo  {journal} {Phys. Rev. Lett.}\ }\textbf {\bibinfo {volume} {121}},\
  \bibinfo {pages} {126401} (\bibinfo {year} {2018}{\natexlab{b}})}\BibitemShut
  {NoStop}%
\bibitem [{\citenamefont {Huang}\ and\ \citenamefont {Liu}(2019)}]{Huang19}%
  \BibitemOpen
  \bibfield  {author} {\bibinfo {author} {\bibfnamefont {H.}~\bibnamefont
  {Huang}}\ and\ \bibinfo {author} {\bibfnamefont {F.}~\bibnamefont {Liu}},\
  }\bibfield  {title} {\bibinfo {title} {Comparison of quantum spin {H}all
  states in quasicrystals and crystals},\ }\href
  {https://doi.org/10.1103/PhysRevB.100.085119} {\bibfield  {journal} {\bibinfo
   {journal} {Phys. Rev. B}\ }\textbf {\bibinfo {volume} {100}},\ \bibinfo
  {pages} {085119} (\bibinfo {year} {2019})}\BibitemShut {NoStop}%
\bibitem [{\citenamefont {He}\ \emph {et~al.}(2019)\citenamefont {He},
  \citenamefont {Ding}, \citenamefont {Zhou}, \citenamefont {Wang},\ and\
  \citenamefont {Gong}}]{He19}%
  \BibitemOpen
  \bibfield  {author} {\bibinfo {author} {\bibfnamefont {A.-L.}\ \bibnamefont
  {He}}, \bibinfo {author} {\bibfnamefont {L.-R.}\ \bibnamefont {Ding}},
  \bibinfo {author} {\bibfnamefont {Y.}~\bibnamefont {Zhou}}, \bibinfo {author}
  {\bibfnamefont {Y.-F.}\ \bibnamefont {Wang}},\ and\ \bibinfo {author}
  {\bibfnamefont {C.-D.}\ \bibnamefont {Gong}},\ }\bibfield  {title} {\bibinfo
  {title} {Quasicrystalline {C}hern insulators},\ }\href
  {https://doi.org/10.1103/PhysRevB.100.214109} {\bibfield  {journal} {\bibinfo
   {journal} {Phys. Rev. B}\ }\textbf {\bibinfo {volume} {100}},\ \bibinfo
  {pages} {214109} (\bibinfo {year} {2019})}\BibitemShut {NoStop}%
\bibitem [{\citenamefont {Varjas}\ \emph {et~al.}(2019)\citenamefont {Varjas},
  \citenamefont {Lau}, \citenamefont {P\"oyh\"onen}, \citenamefont {Akhmerov},
  \citenamefont {Pikulin},\ and\ \citenamefont {Fulga}}]{Varjas19}%
  \BibitemOpen
  \bibfield  {author} {\bibinfo {author} {\bibfnamefont {D.}~\bibnamefont
  {Varjas}}, \bibinfo {author} {\bibfnamefont {A.}~\bibnamefont {Lau}},
  \bibinfo {author} {\bibfnamefont {K.}~\bibnamefont {P\"oyh\"onen}}, \bibinfo
  {author} {\bibfnamefont {A.~R.}\ \bibnamefont {Akhmerov}}, \bibinfo {author}
  {\bibfnamefont {D.~I.}\ \bibnamefont {Pikulin}},\ and\ \bibinfo {author}
  {\bibfnamefont {I.~C.}\ \bibnamefont {Fulga}},\ }\bibfield  {title} {\bibinfo
  {title} {Topological phases without crystalline counterparts},\ }\href
  {https://doi.org/10.1103/PhysRevLett.123.196401} {\bibfield  {journal}
  {\bibinfo  {journal} {Phys. Rev. Lett.}\ }\textbf {\bibinfo {volume} {123}},\
  \bibinfo {pages} {196401} (\bibinfo {year} {2019})}\BibitemShut {NoStop}%
\bibitem [{\citenamefont {Chen}\ \emph {et~al.}(2020)\citenamefont {Chen},
  \citenamefont {Chen}, \citenamefont {Gao}, \citenamefont {Zhou},\ and\
  \citenamefont {Xu}}]{Chen20}%
  \BibitemOpen
  \bibfield  {author} {\bibinfo {author} {\bibfnamefont {R.}~\bibnamefont
  {Chen}}, \bibinfo {author} {\bibfnamefont {C.-Z.}\ \bibnamefont {Chen}},
  \bibinfo {author} {\bibfnamefont {J.-H.}\ \bibnamefont {Gao}}, \bibinfo
  {author} {\bibfnamefont {B.}~\bibnamefont {Zhou}},\ and\ \bibinfo {author}
  {\bibfnamefont {D.-H.}\ \bibnamefont {Xu}},\ }\bibfield  {title} {\bibinfo
  {title} {Higher-order topological insulators in quasicrystals},\ }\href
  {https://doi.org/10.1103/PhysRevLett.124.036803} {\bibfield  {journal}
  {\bibinfo  {journal} {Phys. Rev. Lett.}\ }\textbf {\bibinfo {volume} {124}},\
  \bibinfo {pages} {036803} (\bibinfo {year} {2020})}\BibitemShut {NoStop}%
\bibitem [{\citenamefont {Duncan}\ \emph {et~al.}(2020)\citenamefont {Duncan},
  \citenamefont {Manna},\ and\ \citenamefont {Nielsen}}]{Duncan20}%
  \BibitemOpen
  \bibfield  {author} {\bibinfo {author} {\bibfnamefont {C.~W.}\ \bibnamefont
  {Duncan}}, \bibinfo {author} {\bibfnamefont {S.}~\bibnamefont {Manna}},\ and\
  \bibinfo {author} {\bibfnamefont {A.~E.~B.}\ \bibnamefont {Nielsen}},\
  }\bibfield  {title} {\bibinfo {title} {Topological models in rotationally
  symmetric quasicrystals},\ }\href
  {https://doi.org/10.1103/PhysRevB.101.115413} {\bibfield  {journal} {\bibinfo
   {journal} {Phys. Rev. B}\ }\textbf {\bibinfo {volume} {101}},\ \bibinfo
  {pages} {115413} (\bibinfo {year} {2020})}\BibitemShut {NoStop}%
\bibitem [{\citenamefont {Hua}\ \emph {et~al.}(2020)\citenamefont {Hua},
  \citenamefont {Chen}, \citenamefont {Zhou},\ and\ \citenamefont
  {Xu}}]{Hua20}%
  \BibitemOpen
  \bibfield  {author} {\bibinfo {author} {\bibfnamefont {C.-B.}\ \bibnamefont
  {Hua}}, \bibinfo {author} {\bibfnamefont {R.}~\bibnamefont {Chen}}, \bibinfo
  {author} {\bibfnamefont {B.}~\bibnamefont {Zhou}},\ and\ \bibinfo {author}
  {\bibfnamefont {D.-H.}\ \bibnamefont {Xu}},\ }\bibfield  {title} {\bibinfo
  {title} {Higher-order topological insulator in a dodecagonal quasicrystal},\
  }\href {https://doi.org/10.1103/PhysRevB.102.241102} {\bibfield  {journal}
  {\bibinfo  {journal} {Phys. Rev. B}\ }\textbf {\bibinfo {volume} {102}},\
  \bibinfo {pages} {241102(R)} (\bibinfo {year} {2020})}\BibitemShut {NoStop}%
\bibitem [{\citenamefont {Kraus}\ \emph {et~al.}(2012)\citenamefont {Kraus},
  \citenamefont {Lahini}, \citenamefont {Ringel}, \citenamefont {Verbin},\ and\
  \citenamefont {Zilberberg}}]{Kraus12}%
  \BibitemOpen
  \bibfield  {author} {\bibinfo {author} {\bibfnamefont {Y.~E.}\ \bibnamefont
  {Kraus}}, \bibinfo {author} {\bibfnamefont {Y.}~\bibnamefont {Lahini}},
  \bibinfo {author} {\bibfnamefont {Z.}~\bibnamefont {Ringel}}, \bibinfo
  {author} {\bibfnamefont {M.}~\bibnamefont {Verbin}},\ and\ \bibinfo {author}
  {\bibfnamefont {O.}~\bibnamefont {Zilberberg}},\ }\bibfield  {title}
  {\bibinfo {title} {Topological states and adiabatic pumping in
  quasicrystals},\ }\href {https://doi.org/10.1103/PhysRevLett.109.106402}
  {\bibfield  {journal} {\bibinfo  {journal} {Phys. Rev. Lett.}\ }\textbf
  {\bibinfo {volume} {109}},\ \bibinfo {pages} {106402} (\bibinfo {year}
  {2012})}\BibitemShut {NoStop}%
\bibitem [{\citenamefont {Nakajima}\ \emph {et~al.}(2021)\citenamefont
  {Nakajima}, \citenamefont {Takei}, \citenamefont {Sakuma}, \citenamefont
  {Kuno}, \citenamefont {Marra},\ and\ \citenamefont {Takahashi}}]{Nakajima21}%
  \BibitemOpen
  \bibfield  {author} {\bibinfo {author} {\bibfnamefont {S.}~\bibnamefont
  {Nakajima}}, \bibinfo {author} {\bibfnamefont {N.}~\bibnamefont {Takei}},
  \bibinfo {author} {\bibfnamefont {K.}~\bibnamefont {Sakuma}}, \bibinfo
  {author} {\bibfnamefont {Y.}~\bibnamefont {Kuno}}, \bibinfo {author}
  {\bibfnamefont {P.}~\bibnamefont {Marra}},\ and\ \bibinfo {author}
  {\bibfnamefont {Y.}~\bibnamefont {Takahashi}},\ }\bibfield  {title} {\bibinfo
  {title} {Competition and interplay between topology and quasi-periodic
  disorder in {T}houless pumping of ultracold atoms},\ }\href
  {https://doi.org/10.1038/s41567-021-01229-9} {\bibfield  {journal} {\bibinfo
  {journal} {Nat. Phys.}\ }\textbf {\bibinfo {volume} {17}},\ \bibinfo {pages}
  {844} (\bibinfo {year} {2021})}\BibitemShut {NoStop}%
\bibitem [{\citenamefont {Yoshii}\ \emph {et~al.}(2021)\citenamefont {Yoshii},
  \citenamefont {Kitamura},\ and\ \citenamefont {Morimoto}}]{Yoshii_2021}%
  \BibitemOpen
  \bibfield  {author} {\bibinfo {author} {\bibfnamefont {M.}~\bibnamefont
  {Yoshii}}, \bibinfo {author} {\bibfnamefont {S.}~\bibnamefont {Kitamura}},\
  and\ \bibinfo {author} {\bibfnamefont {T.}~\bibnamefont {Morimoto}},\
  }\bibfield  {title} {\bibinfo {title} {Topological charge pumping in
  quasiperiodic systems characterized by the {B}ott index},\ }\href
  {https://doi.org/10.1103/PhysRevB.104.155126} {\bibfield  {journal} {\bibinfo
   {journal} {Phys. Rev. B}\ }\textbf {\bibinfo {volume} {104}},\ \bibinfo
  {pages} {155126} (\bibinfo {year} {2021})}\BibitemShut {NoStop}%
\bibitem [{\citenamefont {Koshino}\ and\ \citenamefont
  {Oka}(2022)}]{Koshino22}%
  \BibitemOpen
  \bibfield  {author} {\bibinfo {author} {\bibfnamefont {M.}~\bibnamefont
  {Koshino}}\ and\ \bibinfo {author} {\bibfnamefont {H.}~\bibnamefont {Oka}},\
  }\bibfield  {title} {\bibinfo {title} {Topological invariants in
  two-dimensional quasicrystals},\ }\href
  {https://doi.org/10.1103/PhysRevResearch.4.013028} {\bibfield  {journal}
  {\bibinfo  {journal} {Phys. Rev. Res.}\ }\textbf {\bibinfo {volume} {4}},\
  \bibinfo {pages} {013028} (\bibinfo {year} {2022})}\BibitemShut {NoStop}%
\bibitem [{\citenamefont {Yamamoto}\ and\ \citenamefont
  {Koshino}(2022)}]{Yamamoto22}%
  \BibitemOpen
  \bibfield  {author} {\bibinfo {author} {\bibfnamefont {K.}~\bibnamefont
  {Yamamoto}}\ and\ \bibinfo {author} {\bibfnamefont {M.}~\bibnamefont
  {Koshino}},\ }\bibfield  {title} {\bibinfo {title} {Topological gap labeling
  with third {C}hern numbers in three-dimensional quasicrystals},\ }\href
  {https://doi.org/10.1103/PhysRevB.105.115410} {\bibfield  {journal} {\bibinfo
   {journal} {Phys. Rev. B}\ }\textbf {\bibinfo {volume} {105}},\ \bibinfo
  {pages} {115410} (\bibinfo {year} {2022})}\BibitemShut {NoStop}%
\bibitem [{\citenamefont {Timusk}\ \emph {et~al.}(2013)\citenamefont {Timusk},
  \citenamefont {Carbotte}, \citenamefont {Homes}, \citenamefont {Basov},\ and\
  \citenamefont {Sharapov}}]{Timusk13}%
  \BibitemOpen
  \bibfield  {author} {\bibinfo {author} {\bibfnamefont {T.}~\bibnamefont
  {Timusk}}, \bibinfo {author} {\bibfnamefont {J.~P.}\ \bibnamefont
  {Carbotte}}, \bibinfo {author} {\bibfnamefont {C.~C.}\ \bibnamefont {Homes}},
  \bibinfo {author} {\bibfnamefont {D.~N.}\ \bibnamefont {Basov}},\ and\
  \bibinfo {author} {\bibfnamefont {S.~G.}\ \bibnamefont {Sharapov}},\
  }\bibfield  {title} {\bibinfo {title} {Three-dimensional {D}irac fermions in
  quasicrystals as seen via optical conductivity},\ }\href
  {https://doi.org/10.1103/PhysRevB.87.235121} {\bibfield  {journal} {\bibinfo
  {journal} {Phys. Rev. B}\ }\textbf {\bibinfo {volume} {87}},\ \bibinfo
  {pages} {235121} (\bibinfo {year} {2013})}\BibitemShut {NoStop}%
\bibitem [{\citenamefont {Pixley}\ \emph {et~al.}(2018)\citenamefont {Pixley},
  \citenamefont {Wilson}, \citenamefont {Huse},\ and\ \citenamefont
  {Gopalakrishnan}}]{Pixley18}%
  \BibitemOpen
  \bibfield  {author} {\bibinfo {author} {\bibfnamefont {J.~H.}\ \bibnamefont
  {Pixley}}, \bibinfo {author} {\bibfnamefont {J.~H.}\ \bibnamefont {Wilson}},
  \bibinfo {author} {\bibfnamefont {D.~A.}\ \bibnamefont {Huse}},\ and\
  \bibinfo {author} {\bibfnamefont {S.}~\bibnamefont {Gopalakrishnan}},\
  }\bibfield  {title} {\bibinfo {title} {Weyl semimetal to metal phase
  transitions driven by quasiperiodic potentials},\ }\href
  {https://doi.org/10.1103/PhysRevLett.120.207604} {\bibfield  {journal}
  {\bibinfo  {journal} {Phys. Rev. Lett.}\ }\textbf {\bibinfo {volume} {120}},\
  \bibinfo {pages} {207604} (\bibinfo {year} {2018})}\BibitemShut {NoStop}%
\bibitem [{\citenamefont {Mastropietro}(2020)}]{Mastropietro20}%
  \BibitemOpen
  \bibfield  {author} {\bibinfo {author} {\bibfnamefont {V.}~\bibnamefont
  {Mastropietro}},\ }\bibfield  {title} {\bibinfo {title} {Stability of {W}eyl
  semimetals with quasiperiodic disorder},\ }\href
  {https://doi.org/10.1103/PhysRevB.102.045101} {\bibfield  {journal} {\bibinfo
   {journal} {Phys. Rev. B}\ }\textbf {\bibinfo {volume} {102}},\ \bibinfo
  {pages} {045101} (\bibinfo {year} {2020})}\BibitemShut {NoStop}%
\bibitem [{\citenamefont {Cain}\ \emph {et~al.}(2020)\citenamefont {Cain},
  \citenamefont {Azizi}, \citenamefont {Conrad}, \citenamefont {Griffin},\ and\
  \citenamefont {Zettl}}]{JeffreyPNAS20}%
  \BibitemOpen
  \bibfield  {author} {\bibinfo {author} {\bibfnamefont {J.~D.}\ \bibnamefont
  {Cain}}, \bibinfo {author} {\bibfnamefont {A.}~\bibnamefont {Azizi}},
  \bibinfo {author} {\bibfnamefont {M.}~\bibnamefont {Conrad}}, \bibinfo
  {author} {\bibfnamefont {S.~M.}\ \bibnamefont {Griffin}},\ and\ \bibinfo
  {author} {\bibfnamefont {A.}~\bibnamefont {Zettl}},\ }\bibfield  {title}
  {\bibinfo {title} {Layer-dependent topological phase in a two-dimensional
  quasicrystal and approximant},\ }\href
  {https://doi.org/10.1073/pnas.2015164117} {\bibfield  {journal} {\bibinfo
  {journal} {Proc. Natl. Acad. Sci. USA}\ }\textbf {\bibinfo {volume} {117}},\
  \bibinfo {pages} {26135} (\bibinfo {year} {2020})}\BibitemShut {NoStop}%
\bibitem [{\citenamefont {Grossi~e Fonseca}\ \emph {et~al.}(2023)\citenamefont
  {Grossi~e Fonseca}, \citenamefont {Christensen}, \citenamefont
  {Joannopoulos},\ and\ \citenamefont {Solja\ifmmode \check{c}\else
  \v{c}\fi{}i\ifmmode~\acute{c}\else \'{c}\fi{}}}]{Grossi23}%
  \BibitemOpen
  \bibfield  {author} {\bibinfo {author} {\bibfnamefont {A.}~\bibnamefont
  {Grossi~e Fonseca}}, \bibinfo {author} {\bibfnamefont {T.}~\bibnamefont
  {Christensen}}, \bibinfo {author} {\bibfnamefont {J.~D.}\ \bibnamefont
  {Joannopoulos}},\ and\ \bibinfo {author} {\bibfnamefont {M.}~\bibnamefont
  {Solja\ifmmode \check{c}\else \v{c}\fi{}i\ifmmode~\acute{c}\else
  \'{c}\fi{}}},\ }\bibfield  {title} {\bibinfo {title} {Quasicrystalline {W}eyl
  points and dense {F}ermi-{B}ragg arcs},\ }\href
  {https://doi.org/10.1103/PhysRevB.108.L121109} {\bibfield  {journal}
  {\bibinfo  {journal} {Phys. Rev. B}\ }\textbf {\bibinfo {volume} {108}},\
  \bibinfo {pages} {L121109} (\bibinfo {year} {2023})}\BibitemShut {NoStop}%
\bibitem [{\citenamefont {Mao}\ \emph {et~al.}(2024)\citenamefont {Mao},
  \citenamefont {Tao}, \citenamefont {Wang}, \citenamefont {Zeng},\ and\
  \citenamefont {Xu}}]{Mao23arXiv}%
  \BibitemOpen
  \bibfield  {author} {\bibinfo {author} {\bibfnamefont {Y.-F.}\ \bibnamefont
  {Mao}}, \bibinfo {author} {\bibfnamefont {Y.-L.}\ \bibnamefont {Tao}},
  \bibinfo {author} {\bibfnamefont {J.-H.}\ \bibnamefont {Wang}}, \bibinfo
  {author} {\bibfnamefont {Q.-B.}\ \bibnamefont {Zeng}},\ and\ \bibinfo
  {author} {\bibfnamefont {Y.}~\bibnamefont {Xu}},\ }\bibfield  {title}
  {\bibinfo {title} {Higher-order topological insulators in three dimensions
  without crystalline counterparts},\ }\bibfield  {journal} {\bibinfo
  {journal} {Phys. Rev. B}\ }\textbf {\bibinfo {volume} {109}},\ \href
  {https://doi.org/10.1103/PhysRevB.109.134205} {10.1103/PhysRevB.109.134205}
  (\bibinfo {year} {2024})\BibitemShut {NoStop}%
\bibitem [{\citenamefont {Chen}\ \emph {et~al.}(2023)\citenamefont {Chen},
  \citenamefont {Zhou},\ and\ \citenamefont {Xu}}]{Chen23arXiv}%
  \BibitemOpen
  \bibfield  {author} {\bibinfo {author} {\bibfnamefont {R.}~\bibnamefont
  {Chen}}, \bibinfo {author} {\bibfnamefont {B.}~\bibnamefont {Zhou}},\ and\
  \bibinfo {author} {\bibfnamefont {D.-H.}\ \bibnamefont {Xu}},\ }\bibfield
  {title} {\bibinfo {title} {Quasicrystalline second-order topological
  semimetals},\ }\href {https://doi.org/10.1103/PhysRevB.108.195306} {\bibfield
   {journal} {\bibinfo  {journal} {Phys. Rev. B}\ }\textbf {\bibinfo {volume}
  {108}},\ \bibinfo {pages} {195306} (\bibinfo {year} {2023})}\BibitemShut
  {NoStop}%
\bibitem [{\citenamefont {Meng}\ and\ \citenamefont {Balents}(2012)}]{Meng12}%
  \BibitemOpen
  \bibfield  {author} {\bibinfo {author} {\bibfnamefont {T.}~\bibnamefont
  {Meng}}\ and\ \bibinfo {author} {\bibfnamefont {L.}~\bibnamefont {Balents}},\
  }\bibfield  {title} {\bibinfo {title} {Weyl superconductors},\ }\href
  {https://doi.org/10.1103/PhysRevB.86.054504} {\bibfield  {journal} {\bibinfo
  {journal} {Phys. Rev. B}\ }\textbf {\bibinfo {volume} {86}},\ \bibinfo
  {pages} {054504} (\bibinfo {year} {2012})}\BibitemShut {NoStop}%
\bibitem [{\citenamefont {Sau}\ and\ \citenamefont {Tewari}(2012)}]{Sau12}%
  \BibitemOpen
  \bibfield  {author} {\bibinfo {author} {\bibfnamefont {J.~D.}\ \bibnamefont
  {Sau}}\ and\ \bibinfo {author} {\bibfnamefont {S.}~\bibnamefont {Tewari}},\
  }\bibfield  {title} {\bibinfo {title} {Topologically protected surface
  {M}ajorana arcs and bulk {W}eyl fermions in ferromagnetic superconductors},\
  }\href {https://doi.org/10.1103/PhysRevB.86.104509} {\bibfield  {journal}
  {\bibinfo  {journal} {Phys. Rev. B}\ }\textbf {\bibinfo {volume} {86}},\
  \bibinfo {pages} {104509} (\bibinfo {year} {2012})}\BibitemShut {NoStop}%
\bibitem [{\citenamefont {Fischer}\ \emph {et~al.}(2014)\citenamefont
  {Fischer}, \citenamefont {Neupert}, \citenamefont {Platt}, \citenamefont
  {Schnyder}, \citenamefont {Hanke}, \citenamefont {Goryo}, \citenamefont
  {Thomale},\ and\ \citenamefont {Sigrist}}]{Fischer14}%
  \BibitemOpen
  \bibfield  {author} {\bibinfo {author} {\bibfnamefont {M.~H.}\ \bibnamefont
  {Fischer}}, \bibinfo {author} {\bibfnamefont {T.}~\bibnamefont {Neupert}},
  \bibinfo {author} {\bibfnamefont {C.}~\bibnamefont {Platt}}, \bibinfo
  {author} {\bibfnamefont {A.~P.}\ \bibnamefont {Schnyder}}, \bibinfo {author}
  {\bibfnamefont {W.}~\bibnamefont {Hanke}}, \bibinfo {author} {\bibfnamefont
  {J.}~\bibnamefont {Goryo}}, \bibinfo {author} {\bibfnamefont
  {R.}~\bibnamefont {Thomale}},\ and\ \bibinfo {author} {\bibfnamefont
  {M.}~\bibnamefont {Sigrist}},\ }\bibfield  {title} {\bibinfo {title} {Chiral
  $d$-wave superconductivity in {S}r{P}t{A}s},\ }\href
  {https://doi.org/10.1103/PhysRevB.89.020509} {\bibfield  {journal} {\bibinfo
  {journal} {Phys. Rev. B}\ }\textbf {\bibinfo {volume} {89}},\ \bibinfo
  {pages} {020509(R)} (\bibinfo {year} {2014})}\BibitemShut {NoStop}%
\bibitem [{\citenamefont {Yang}\ \emph {et~al.}(2014)\citenamefont {Yang},
  \citenamefont {Pan},\ and\ \citenamefont {Zhang}}]{Yang14}%
  \BibitemOpen
  \bibfield  {author} {\bibinfo {author} {\bibfnamefont {S.~A.}\ \bibnamefont
  {Yang}}, \bibinfo {author} {\bibfnamefont {H.}~\bibnamefont {Pan}},\ and\
  \bibinfo {author} {\bibfnamefont {F.}~\bibnamefont {Zhang}},\ }\bibfield
  {title} {\bibinfo {title} {Dirac and {W}eyl superconductors in three
  dimensions},\ }\href {https://doi.org/10.1103/PhysRevLett.113.046401}
  {\bibfield  {journal} {\bibinfo  {journal} {Phys. Rev. Lett.}\ }\textbf
  {\bibinfo {volume} {113}},\ \bibinfo {pages} {046401} (\bibinfo {year}
  {2014})}\BibitemShut {NoStop}%
\bibitem [{\citenamefont {Bednik}\ \emph {et~al.}(2015)\citenamefont {Bednik},
  \citenamefont {Zyuzin},\ and\ \citenamefont {Burkov}}]{Bendik_2015}%
  \BibitemOpen
  \bibfield  {author} {\bibinfo {author} {\bibfnamefont {G.}~\bibnamefont
  {Bednik}}, \bibinfo {author} {\bibfnamefont {A.~A.}\ \bibnamefont {Zyuzin}},\
  and\ \bibinfo {author} {\bibfnamefont {A.~A.}\ \bibnamefont {Burkov}},\
  }\bibfield  {title} {\bibinfo {title} {Superconductivity in {W}eyl metals},\
  }\href {https://doi.org/10.1103/PhysRevB.92.035153} {\bibfield  {journal}
  {\bibinfo  {journal} {Phys. Rev. B}\ }\textbf {\bibinfo {volume} {92}},\
  \bibinfo {pages} {035153} (\bibinfo {year} {2015})}\BibitemShut {NoStop}%
\bibitem [{\citenamefont {Yuan}\ \emph {et~al.}(2017)\citenamefont {Yuan},
  \citenamefont {He},\ and\ \citenamefont {Law}}]{Yuan17}%
  \BibitemOpen
  \bibfield  {author} {\bibinfo {author} {\bibfnamefont {N.~F.~Q.}\
  \bibnamefont {Yuan}}, \bibinfo {author} {\bibfnamefont {W.-Y.}\ \bibnamefont
  {He}},\ and\ \bibinfo {author} {\bibfnamefont {K.~T.}\ \bibnamefont {Law}},\
  }\bibfield  {title} {\bibinfo {title} {Superconductivity-induced
  ferromagnetism and {W}eyl superconductivity in {N}b-doped {B}i$_2${S}e$_3$},\
  }\href {https://doi.org/10.1103/PhysRevB.95.201109} {\bibfield  {journal}
  {\bibinfo  {journal} {Phys. Rev. B}\ }\textbf {\bibinfo {volume} {95}},\
  \bibinfo {pages} {201109(R)} (\bibinfo {year} {2017})}\BibitemShut {NoStop}%
\bibitem [{\citenamefont {Li}\ and\ \citenamefont {Haldane}(2018)}]{Li_2018}%
  \BibitemOpen
  \bibfield  {author} {\bibinfo {author} {\bibfnamefont {Y.}~\bibnamefont
  {Li}}\ and\ \bibinfo {author} {\bibfnamefont {F.~D.~M.}\ \bibnamefont
  {Haldane}},\ }\bibfield  {title} {\bibinfo {title} {Topological nodal
  {C}ooper pairing in doped {W}eyl metals},\ }\href
  {https://doi.org/10.1103/PhysRevLett.120.067003} {\bibfield  {journal}
  {\bibinfo  {journal} {Phys. Rev. Lett.}\ }\textbf {\bibinfo {volume} {120}},\
  \bibinfo {pages} {067003} (\bibinfo {year} {2018})}\BibitemShut {NoStop}%
\bibitem [{\citenamefont {Okugawa}\ and\ \citenamefont
  {Yokoyama}(2018)}]{Okugawa18}%
  \BibitemOpen
  \bibfield  {author} {\bibinfo {author} {\bibfnamefont {R.}~\bibnamefont
  {Okugawa}}\ and\ \bibinfo {author} {\bibfnamefont {T.}~\bibnamefont
  {Yokoyama}},\ }\bibfield  {title} {\bibinfo {title} {Generic phase diagram
  for {W}eyl superconductivity in mirror-symmetric superconductors},\ }\href
  {https://doi.org/10.1103/PhysRevB.97.060504} {\bibfield  {journal} {\bibinfo
  {journal} {Phys. Rev. B}\ }\textbf {\bibinfo {volume} {97}},\ \bibinfo
  {pages} {060504(R)} (\bibinfo {year} {2018})}\BibitemShut {NoStop}%
\bibitem [{\citenamefont {Sumita}\ and\ \citenamefont
  {Yanase}(2018)}]{Sumita18}%
  \BibitemOpen
  \bibfield  {author} {\bibinfo {author} {\bibfnamefont {S.}~\bibnamefont
  {Sumita}}\ and\ \bibinfo {author} {\bibfnamefont {Y.}~\bibnamefont
  {Yanase}},\ }\bibfield  {title} {\bibinfo {title} {Unconventional
  superconducting gap structure protected by space group symmetry},\ }\href
  {https://doi.org/10.1103/PhysRevB.97.134512} {\bibfield  {journal} {\bibinfo
  {journal} {Phys. Rev. B}\ }\textbf {\bibinfo {volume} {97}},\ \bibinfo
  {pages} {134512} (\bibinfo {year} {2018})}\BibitemShut {NoStop}%
\bibitem [{\citenamefont {Nakai}\ and\ \citenamefont {Nomura}(2020)}]{Nakai20}%
  \BibitemOpen
  \bibfield  {author} {\bibinfo {author} {\bibfnamefont {R.}~\bibnamefont
  {Nakai}}\ and\ \bibinfo {author} {\bibfnamefont {K.}~\bibnamefont {Nomura}},\
  }\bibfield  {title} {\bibinfo {title} {Weyl superconductor phases in a
  {W}eyl-semimetal/superconductor multilayer},\ }\href
  {https://doi.org/10.1103/PhysRevB.101.094510} {\bibfield  {journal} {\bibinfo
   {journal} {Phys. Rev. B}\ }\textbf {\bibinfo {volume} {101}},\ \bibinfo
  {pages} {094510} (\bibinfo {year} {2020})}\BibitemShut {NoStop}%
\bibitem [{\citenamefont {Wan}\ \emph {et~al.}(2011)\citenamefont {Wan},
  \citenamefont {Turner}, \citenamefont {Vishwanath},\ and\ \citenamefont
  {Savrasov}}]{Wan11}%
  \BibitemOpen
  \bibfield  {author} {\bibinfo {author} {\bibfnamefont {X.}~\bibnamefont
  {Wan}}, \bibinfo {author} {\bibfnamefont {A.~M.}\ \bibnamefont {Turner}},
  \bibinfo {author} {\bibfnamefont {A.}~\bibnamefont {Vishwanath}},\ and\
  \bibinfo {author} {\bibfnamefont {S.~Y.}\ \bibnamefont {Savrasov}},\
  }\bibfield  {title} {\bibinfo {title} {Topological semimetal and {F}ermi-arc
  surface states in the electronic structure of pyrochlore iridates},\ }\href
  {https://doi.org/10.1103/PhysRevB.83.205101} {\bibfield  {journal} {\bibinfo
  {journal} {Phys. Rev. B}\ }\textbf {\bibinfo {volume} {83}},\ \bibinfo
  {pages} {205101} (\bibinfo {year} {2011})}\BibitemShut {NoStop}%
\bibitem [{\citenamefont {Burkov}\ and\ \citenamefont
  {Balents}(2011)}]{Burkov11}%
  \BibitemOpen
  \bibfield  {author} {\bibinfo {author} {\bibfnamefont {A.~A.}\ \bibnamefont
  {Burkov}}\ and\ \bibinfo {author} {\bibfnamefont {L.}~\bibnamefont
  {Balents}},\ }\bibfield  {title} {\bibinfo {title} {Weyl semimetal in a
  topological insulator multilayer},\ }\href
  {https://doi.org/10.1103/PhysRevLett.107.127205} {\bibfield  {journal}
  {\bibinfo  {journal} {Phys. Rev. Lett.}\ }\textbf {\bibinfo {volume} {107}},\
  \bibinfo {pages} {127205} (\bibinfo {year} {2011})}\BibitemShut {NoStop}%
\bibitem [{\citenamefont {Okugawa}\ and\ \citenamefont
  {Murakami}(2014)}]{Okugawa14}%
  \BibitemOpen
  \bibfield  {author} {\bibinfo {author} {\bibfnamefont {R.}~\bibnamefont
  {Okugawa}}\ and\ \bibinfo {author} {\bibfnamefont {S.}~\bibnamefont
  {Murakami}},\ }\bibfield  {title} {\bibinfo {title} {Dispersion of {F}ermi
  arcs in {W}eyl semimetals and their evolutions to {D}irac cones},\ }\href
  {https://doi.org/10.1103/PhysRevB.89.235315} {\bibfield  {journal} {\bibinfo
  {journal} {Phys. Rev. B}\ }\textbf {\bibinfo {volume} {89}},\ \bibinfo
  {pages} {235315} (\bibinfo {year} {2014})}\BibitemShut {NoStop}%
\bibitem [{\citenamefont {Schnyder}\ and\ \citenamefont
  {Brydon}(2015)}]{Schnyder15}%
  \BibitemOpen
  \bibfield  {author} {\bibinfo {author} {\bibfnamefont {A.~P.}\ \bibnamefont
  {Schnyder}}\ and\ \bibinfo {author} {\bibfnamefont {P.~M.~R.}\ \bibnamefont
  {Brydon}},\ }\bibfield  {title} {\bibinfo {title} {Topological surface states
  in nodal superconductors},\ }\href
  {https://doi.org/10.1088/0953-8984/27/24/243201} {\bibfield  {journal}
  {\bibinfo  {journal} {J. Phys.: Condense. Matter}\ }\textbf {\bibinfo
  {volume} {27}},\ \bibinfo {pages} {243201} (\bibinfo {year}
  {2015})}\BibitemShut {NoStop}%
\bibitem [{\citenamefont {Read}\ and\ \citenamefont {Green}(2000)}]{Read_2000}%
  \BibitemOpen
  \bibfield  {author} {\bibinfo {author} {\bibfnamefont {N.}~\bibnamefont
  {Read}}\ and\ \bibinfo {author} {\bibfnamefont {D.}~\bibnamefont {Green}},\
  }\bibfield  {title} {\bibinfo {title} {Paired states of fermions in two
  dimensions with breaking of parity and time-reversal symmetries and the
  fractional quantum {H}all effect},\ }\href
  {https://doi.org/10.1103/PhysRevB.61.10267} {\bibfield  {journal} {\bibinfo
  {journal} {Phys. Rev. B}\ }\textbf {\bibinfo {volume} {61}},\ \bibinfo
  {pages} {10267} (\bibinfo {year} {2000})}\BibitemShut {NoStop}%
\bibitem [{\citenamefont {Fulga}\ \emph {et~al.}(2016)\citenamefont {Fulga},
  \citenamefont {Pikulin},\ and\ \citenamefont {Loring}}]{Fulga16}%
  \BibitemOpen
  \bibfield  {author} {\bibinfo {author} {\bibfnamefont {I.~C.}\ \bibnamefont
  {Fulga}}, \bibinfo {author} {\bibfnamefont {D.~I.}\ \bibnamefont {Pikulin}},\
  and\ \bibinfo {author} {\bibfnamefont {T.~A.}\ \bibnamefont {Loring}},\
  }\bibfield  {title} {\bibinfo {title} {Aperiodic weak topological
  superconductors},\ }\href {https://doi.org/10.1103/PhysRevLett.116.257002}
  {\bibfield  {journal} {\bibinfo  {journal} {Phys. Rev. Lett.}\ }\textbf
  {\bibinfo {volume} {116}},\ \bibinfo {pages} {257002} (\bibinfo {year}
  {2016})}\BibitemShut {NoStop}%
\bibitem [{\citenamefont {Cao}\ \emph {et~al.}(2020)\citenamefont {Cao},
  \citenamefont {Zhang}, \citenamefont {Liu}, \citenamefont {Liu},
  \citenamefont {Chen},\ and\ \citenamefont {Yang}}]{Cao20}%
  \BibitemOpen
  \bibfield  {author} {\bibinfo {author} {\bibfnamefont {Y.}~\bibnamefont
  {Cao}}, \bibinfo {author} {\bibfnamefont {Y.}~\bibnamefont {Zhang}}, \bibinfo
  {author} {\bibfnamefont {Y.-B.}\ \bibnamefont {Liu}}, \bibinfo {author}
  {\bibfnamefont {C.-C.}\ \bibnamefont {Liu}}, \bibinfo {author} {\bibfnamefont
  {W.-Q.}\ \bibnamefont {Chen}},\ and\ \bibinfo {author} {\bibfnamefont
  {F.}~\bibnamefont {Yang}},\ }\bibfield  {title} {\bibinfo {title}
  {Kohn-{L}uttinger mechanism driven exotic topological superconductivity on
  the {P}enrose lattice},\ }\href
  {https://doi.org/10.1103/PhysRevLett.125.017002} {\bibfield  {journal}
  {\bibinfo  {journal} {Phys. Rev. Lett.}\ }\textbf {\bibinfo {volume} {125}},\
  \bibinfo {pages} {017002} (\bibinfo {year} {2020})}\BibitemShut {NoStop}%
\bibitem [{\citenamefont {Wang}\ \emph {et~al.}(2022)\citenamefont {Wang},
  \citenamefont {Liu},\ and\ \citenamefont {Huang}}]{Wang22}%
  \BibitemOpen
  \bibfield  {author} {\bibinfo {author} {\bibfnamefont {C.}~\bibnamefont
  {Wang}}, \bibinfo {author} {\bibfnamefont {F.}~\bibnamefont {Liu}},\ and\
  \bibinfo {author} {\bibfnamefont {H.}~\bibnamefont {Huang}},\ }\bibfield
  {title} {\bibinfo {title} {Effective model for fractional topological corner
  modes in quasicrystals},\ }\href
  {https://doi.org/10.1103/PhysRevLett.129.056403} {\bibfield  {journal}
  {\bibinfo  {journal} {Phys. Rev. Lett.}\ }\textbf {\bibinfo {volume} {129}},\
  \bibinfo {pages} {056403} (\bibinfo {year} {2022})}\BibitemShut {NoStop}%
\bibitem [{\citenamefont {Liu}\ \emph {et~al.}(2023)\citenamefont {Liu},
  \citenamefont {Zhang}, \citenamefont {Chen},\ and\ \citenamefont
  {Yang}}]{Liu23}%
  \BibitemOpen
  \bibfield  {author} {\bibinfo {author} {\bibfnamefont {Y.-B.}\ \bibnamefont
  {Liu}}, \bibinfo {author} {\bibfnamefont {Y.}~\bibnamefont {Zhang}}, \bibinfo
  {author} {\bibfnamefont {W.-Q.}\ \bibnamefont {Chen}},\ and\ \bibinfo
  {author} {\bibfnamefont {F.}~\bibnamefont {Yang}},\ }\bibfield  {title}
  {\bibinfo {title} {High-angular-momentum topological superconductivities in
  twisted bilayer quasicrystal systems},\ }\href
  {https://doi.org/10.1103/PhysRevB.107.014501} {\bibfield  {journal} {\bibinfo
   {journal} {Phys. Rev. B}\ }\textbf {\bibinfo {volume} {107}},\ \bibinfo
  {pages} {014501} (\bibinfo {year} {2023})}\BibitemShut {NoStop}%
\bibitem [{\citenamefont {Ghadimi}\ \emph {et~al.}(2021)\citenamefont
  {Ghadimi}, \citenamefont {Sugimoto}, \citenamefont {Tanaka},\ and\
  \citenamefont {Tohyama}}]{Ghadimi21}%
  \BibitemOpen
  \bibfield  {author} {\bibinfo {author} {\bibfnamefont {R.}~\bibnamefont
  {Ghadimi}}, \bibinfo {author} {\bibfnamefont {T.}~\bibnamefont {Sugimoto}},
  \bibinfo {author} {\bibfnamefont {K.}~\bibnamefont {Tanaka}},\ and\ \bibinfo
  {author} {\bibfnamefont {T.}~\bibnamefont {Tohyama}},\ }\bibfield  {title}
  {\bibinfo {title} {Topological superconductivity in quasicrystals},\ }\href
  {https://doi.org/10.1103/PhysRevB.104.144511} {\bibfield  {journal} {\bibinfo
   {journal} {Phys. Rev. B}\ }\textbf {\bibinfo {volume} {104}},\ \bibinfo
  {pages} {144511} (\bibinfo {year} {2021})}\BibitemShut {NoStop}%
\bibitem [{\citenamefont {Hori}\ \emph {et~al.}()\citenamefont {Hori},
  \citenamefont {Sugimoto}, \citenamefont {Tohyama},\ and\ \citenamefont
  {Tanaka}}]{Hori24}%
  \BibitemOpen
  \bibfield  {author} {\bibinfo {author} {\bibfnamefont {M.}~\bibnamefont
  {Hori}}, \bibinfo {author} {\bibfnamefont {T.}~\bibnamefont {Sugimoto}},
  \bibinfo {author} {\bibfnamefont {T.}~\bibnamefont {Tohyama}},\ and\ \bibinfo
  {author} {\bibfnamefont {K.}~\bibnamefont {Tanaka}},\ }\bibfield  {title}
  {\bibinfo {title} {Self-consistent study of topological superconductivity in
  two-dimensional quasicrystals},\ }\href
  {https://doi.org/10.48550/arXiv.2401.06355} {\bibinfo  {journal} {arXiv
  preprint arXiv:2401.06355}\ }\BibitemShut {NoStop}%
\bibitem [{\citenamefont {Sato}\ \emph {et~al.}(2009)\citenamefont {Sato},
  \citenamefont {Takahashi},\ and\ \citenamefont {Fujimoto}}]{Sato09}%
  \BibitemOpen
\bibfield  {journal} {  }\bibfield  {author} {\bibinfo {author} {\bibfnamefont
  {M.}~\bibnamefont {Sato}}, \bibinfo {author} {\bibfnamefont {Y.}~\bibnamefont
  {Takahashi}},\ and\ \bibinfo {author} {\bibfnamefont {S.}~\bibnamefont
  {Fujimoto}},\ }\bibfield  {title} {\bibinfo {title} {Non-{A}belian
  topological order in $s$-wave superfluids of ultracold fermionic atoms},\
  }\href {https://doi.org/10.1103/PhysRevLett.103.020401} {\bibfield  {journal}
  {\bibinfo  {journal} {Phys. Rev. Lett.}\ }\textbf {\bibinfo {volume} {103}},\
  \bibinfo {pages} {020401} (\bibinfo {year} {2009})}\BibitemShut {NoStop}%
\bibitem [{\citenamefont {Sato}\ \emph {et~al.}(2010)\citenamefont {Sato},
  \citenamefont {Takahashi},\ and\ \citenamefont {Fujimoto}}]{Sato10}%
  \BibitemOpen
  \bibfield  {author} {\bibinfo {author} {\bibfnamefont {M.}~\bibnamefont
  {Sato}}, \bibinfo {author} {\bibfnamefont {Y.}~\bibnamefont {Takahashi}},\
  and\ \bibinfo {author} {\bibfnamefont {S.}~\bibnamefont {Fujimoto}},\
  }\bibfield  {title} {\bibinfo {title} {Non-{A}belian topological orders and
  {M}ajorana fermions in spin-singlet superconductors},\ }\href
  {https://doi.org/10.1103/PhysRevB.82.134521} {\bibfield  {journal} {\bibinfo
  {journal} {Phys. Rev. B}\ }\textbf {\bibinfo {volume} {82}},\ \bibinfo
  {pages} {134521} (\bibinfo {year} {2010})}\BibitemShut {NoStop}%
\bibitem [{\citenamefont {Altland}\ and\ \citenamefont
  {Zirnbauer}(1997)}]{Altland_1997}%
  \BibitemOpen
  \bibfield  {author} {\bibinfo {author} {\bibfnamefont {A.}~\bibnamefont
  {Altland}}\ and\ \bibinfo {author} {\bibfnamefont {M.~R.}\ \bibnamefont
  {Zirnbauer}},\ }\bibfield  {title} {\bibinfo {title} {Nonstandard symmetry
  classes in mesoscopic normal-superconducting hybrid structures},\ }\href
  {https://doi.org/10.1103/PhysRevB.55.1142} {\bibfield  {journal} {\bibinfo
  {journal} {Phys. Rev. B}\ }\textbf {\bibinfo {volume} {55}},\ \bibinfo
  {pages} {1142} (\bibinfo {year} {1997})}\BibitemShut {NoStop}%
\bibitem [{\citenamefont {Schnyder}\ \emph {et~al.}(2008)\citenamefont
  {Schnyder}, \citenamefont {Ryu}, \citenamefont {Furusaki},\ and\
  \citenamefont {Ludwig}}]{Schnyder_2008}%
  \BibitemOpen
  \bibfield  {author} {\bibinfo {author} {\bibfnamefont {A.~P.}\ \bibnamefont
  {Schnyder}}, \bibinfo {author} {\bibfnamefont {S.}~\bibnamefont {Ryu}},
  \bibinfo {author} {\bibfnamefont {A.}~\bibnamefont {Furusaki}},\ and\
  \bibinfo {author} {\bibfnamefont {A.~W.~W.}\ \bibnamefont {Ludwig}},\
  }\bibfield  {title} {\bibinfo {title} {Classification of topological
  insulators and superconductors in three spatial dimensions},\ }\href
  {https://doi.org/10.1103/PhysRevB.78.195125} {\bibfield  {journal} {\bibinfo
  {journal} {Phys. Rev. B}\ }\textbf {\bibinfo {volume} {78}},\ \bibinfo
  {pages} {195125} (\bibinfo {year} {2008})}\BibitemShut {NoStop}%
\bibitem [{\citenamefont {Chiu}\ \emph {et~al.}(2016)\citenamefont {Chiu},
  \citenamefont {Teo}, \citenamefont {Schnyder},\ and\ \citenamefont
  {Ryu}}]{Chiu_2016}%
  \BibitemOpen
  \bibfield  {author} {\bibinfo {author} {\bibfnamefont {C.-K.}\ \bibnamefont
  {Chiu}}, \bibinfo {author} {\bibfnamefont {J.~C.~Y.}\ \bibnamefont {Teo}},
  \bibinfo {author} {\bibfnamefont {A.~P.}\ \bibnamefont {Schnyder}},\ and\
  \bibinfo {author} {\bibfnamefont {S.}~\bibnamefont {Ryu}},\ }\bibfield
  {title} {\bibinfo {title} {Classification of topological quantum matter with
  symmetries},\ }\href {https://doi.org/10.1103/RevModPhys.88.035005}
  {\bibfield  {journal} {\bibinfo  {journal} {Rev. Mod. Phys.}\ }\textbf
  {\bibinfo {volume} {88}},\ \bibinfo {pages} {035005} (\bibinfo {year}
  {2016})}\BibitemShut {NoStop}%
\bibitem [{\citenamefont {Loring}\ and\ \citenamefont
  {Hastings}(2011)}]{Loring10}%
  \BibitemOpen
  \bibfield  {author} {\bibinfo {author} {\bibfnamefont {T.~A.}\ \bibnamefont
  {Loring}}\ and\ \bibinfo {author} {\bibfnamefont {M.~B.}\ \bibnamefont
  {Hastings}},\ }\bibfield  {title} {\bibinfo {title} {Disordered topological
  insulators via {C}*-algebras},\ }\href
  {https://doi.org/10.1209/0295-5075/92/67004} {\bibfield  {journal} {\bibinfo
  {journal} {Europhys. Lett.}\ }\textbf {\bibinfo {volume} {92}},\ \bibinfo
  {pages} {67004} (\bibinfo {year} {2011})}\BibitemShut {NoStop}%
\bibitem [{\citenamefont {M{\'e}nard}\ \emph {et~al.}(2017)\citenamefont
  {M{\'e}nard}, \citenamefont {Guissart}, \citenamefont {Brun}, \citenamefont
  {Leriche}, \citenamefont {Trif}, \citenamefont {Debontridder}, \citenamefont
  {Demaille}, \citenamefont {Roditchev}, \citenamefont {Simon},\ and\
  \citenamefont {Cren}}]{Menard_2017}%
  \BibitemOpen
  \bibfield  {author} {\bibinfo {author} {\bibfnamefont {G.~C.}\ \bibnamefont
  {M{\'e}nard}}, \bibinfo {author} {\bibfnamefont {S.}~\bibnamefont
  {Guissart}}, \bibinfo {author} {\bibfnamefont {C.}~\bibnamefont {Brun}},
  \bibinfo {author} {\bibfnamefont {R.~T.}\ \bibnamefont {Leriche}}, \bibinfo
  {author} {\bibfnamefont {M.}~\bibnamefont {Trif}}, \bibinfo {author}
  {\bibfnamefont {F.}~\bibnamefont {Debontridder}}, \bibinfo {author}
  {\bibfnamefont {D.}~\bibnamefont {Demaille}}, \bibinfo {author}
  {\bibfnamefont {D.}~\bibnamefont {Roditchev}}, \bibinfo {author}
  {\bibfnamefont {P.}~\bibnamefont {Simon}},\ and\ \bibinfo {author}
  {\bibfnamefont {T.}~\bibnamefont {Cren}},\ }\bibfield  {title} {\bibinfo
  {title} {Two-dimensional topological superconductivity in
  {P}b/{C}o/{S}i(111)},\ }\href {https://doi.org/10.1038/s41467-017-02192-x}
  {\bibfield  {journal} {\bibinfo  {journal} {Nat. Commun.}\ }\textbf {\bibinfo
  {volume} {8}},\ \bibinfo {pages} {2040} (\bibinfo {year} {2017})}\BibitemShut
  {NoStop}%
\bibitem [{\citenamefont {Ara\'ujo}\ and\ \citenamefont
  {Andrade}(2019)}]{Araujo19}%
  \BibitemOpen
  \bibfield  {author} {\bibinfo {author} {\bibfnamefont {R.~N.}\ \bibnamefont
  {Ara\'ujo}}\ and\ \bibinfo {author} {\bibfnamefont {E.~C.}\ \bibnamefont
  {Andrade}},\ }\bibfield  {title} {\bibinfo {title} {Conventional
  superconductivity in quasicrystals},\ }\href
  {https://doi.org/10.1103/PhysRevB.100.014510} {\bibfield  {journal} {\bibinfo
   {journal} {Phys. Rev. B}\ }\textbf {\bibinfo {volume} {100}},\ \bibinfo
  {pages} {014510} (\bibinfo {year} {2019})}\BibitemShut {NoStop}%
\bibitem [{\citenamefont {Janot}(1992)}]{Janot_1992}%
  \BibitemOpen
  \bibfield  {author} {\bibinfo {author} {\bibfnamefont {C.}~\bibnamefont
  {Janot}},\ }\href@noop {} {\emph {\bibinfo {title} {{Quasicrystals - A
  Primer}}}}\ (\bibinfo  {publisher} {Carendon Press, Oxford},\ \bibinfo {year}
  {1992})\BibitemShut {NoStop}%
\bibitem [{\citenamefont {Koga}(2020)}]{Koga_2020}%
  \BibitemOpen
  \bibfield  {author} {\bibinfo {author} {\bibfnamefont {A.}~\bibnamefont
  {Koga}},\ }\bibfield  {title} {\bibinfo {title} {Superlattice structure in
  the antiferromagnetically ordered state in the {H}ubbard model on the
  {A}mmann-{B}eenker tiling},\ }\href
  {https://doi.org/10.1103/PhysRevB.102.115125} {\bibfield  {journal} {\bibinfo
   {journal} {Phys. Rev. B}\ }\textbf {\bibinfo {volume} {102}},\ \bibinfo
  {pages} {115125} (\bibinfo {year} {2020})}\BibitemShut {NoStop}%
\bibitem [{\citenamefont {Conrad}\ \emph {et~al.}(1998)\citenamefont {Conrad},
  \citenamefont {Krumeich},\ and\ \citenamefont {Harbrecht}}]{Conrad_1998}%
  \BibitemOpen
  \bibfield  {author} {\bibinfo {author} {\bibfnamefont {M.}~\bibnamefont
  {Conrad}}, \bibinfo {author} {\bibfnamefont {F.}~\bibnamefont {Krumeich}},\
  and\ \bibinfo {author} {\bibfnamefont {B.}~\bibnamefont {Harbrecht}},\
  }\bibfield  {title} {\bibinfo {title} {A dodecagonal quasicrystalline
  chalcogenide},\ }\href
  {https://doi.org/https://doi.org/10.1002/(SICI)1521-3773(19980605)37:10<1383::AID-ANIE1383>3.0.CO;2-R}
  {\bibfield  {journal} {\bibinfo  {journal} {Angew. Chem. Int. Ed.}\ }\textbf
  {\bibinfo {volume} {37}},\ \bibinfo {pages} {1383} (\bibinfo {year}
  {1998})}\BibitemShut {NoStop}%
\bibitem [{\citenamefont {Gambaudo}\ and\ \citenamefont
  {Vignolo}(2014)}]{Gambaudo_2014}%
  \BibitemOpen
  \bibfield  {author} {\bibinfo {author} {\bibfnamefont {J.-M.}\ \bibnamefont
  {Gambaudo}}\ and\ \bibinfo {author} {\bibfnamefont {P.}~\bibnamefont
  {Vignolo}},\ }\bibfield  {title} {\bibinfo {title} {Brillouin zone labelling
  for quasicrystals},\ }\href {https://doi.org/10.1088/1367-2630/16/4/043013}
  {\bibfield  {journal} {\bibinfo  {journal} {New J. Phys.}\ }\textbf {\bibinfo
  {volume} {16}},\ \bibinfo {pages} {043013} (\bibinfo {year}
  {2014})}\BibitemShut {NoStop}%
\bibitem [{\citenamefont {Koga}\ and\ \citenamefont
  {Tsunetsugu}(2017)}]{Koga_2017}%
  \BibitemOpen
  \bibfield  {author} {\bibinfo {author} {\bibfnamefont {A.}~\bibnamefont
  {Koga}}\ and\ \bibinfo {author} {\bibfnamefont {H.}~\bibnamefont
  {Tsunetsugu}},\ }\bibfield  {title} {\bibinfo {title} {Antiferromagnetic
  order in the {H}ubbard model on the {P}enrose lattice},\ }\href
  {https://doi.org/10.1103/PhysRevB.96.214402} {\bibfield  {journal} {\bibinfo
  {journal} {Phys. Rev. B}\ }\textbf {\bibinfo {volume} {96}},\ \bibinfo
  {pages} {214402} (\bibinfo {year} {2017})}\BibitemShut {NoStop}%
\end{thebibliography}%

\end{document}